\documentclass[12pt,english]{article}
\usepackage[T1]{fontenc}
\usepackage[latin9]{inputenc}
\usepackage{geometry}
\usepackage{babel}
\usepackage{textcomp}
\usepackage{amsmath}
\usepackage{amsthm}
\usepackage{graphicx}
\usepackage{setspace}
\usepackage[authoryear]{natbib}
\usepackage[unicode=true,pdfusetitle,
 bookmarks=true,bookmarksnumbered=false,bookmarksopen=false,
 breaklinks=false,pdfborder={0 0 1},backref=false,colorlinks=false]
 {hyperref}
\hypersetup{
 colorlinks=true,citecolor=blue, urlcolor=cyan, linkcolor=black}
\begin{document}
\date{}
\title{\vskip -2emA frequentist test of proportional colocalization after
selecting relevant genetic variants\thanks{Ashish Patel (\protect\href{mailto:ashish.patel@mrc-bsu.cam.ac.uk}{ashish.patel@mrc-bsu.cam.ac.uk});
John Whittaker (\protect\href{http://john.whittaker@mrc-bsu.cam.ac.uk}{john.whittaker@mrc-bsu.cam.ac.uk});
Stephen Burgess (\protect\href{http://sb452@medschl.cam.ac.uk}{sb452@medschl.cam.ac.uk}).}}
\author{Ashish Patel{\small{} }\textsuperscript{{\small{}a}}{\small{},}
John C.\@ Whittaker{\small{} }\textsuperscript{{\small{}a}}{\small{},}
\& Stephen Burgess{\small{} }\textsuperscript{{\small{}a,b}}}
\maketitle
\begin{center}
{\small{}\vskip -2em}\textsuperscript{{\small{}a}}{\small{} MRC
Biostatistics Unit, University of Cambridge}\\
{\small{}}\textsuperscript{{\small{}b}}{\small{} Cardiovascular
Epidemiology Unit, University of Cambridge}{\small\par}
\par\end{center}
\begin{abstract}
Colocalization analyses assess whether two traits are affected by
the same or distinct causal genetic variants in a single gene region.
A class of Bayesian colocalization tests \citep{Giambartolomei2014,Wallace2021}
are now routinely used in practice; for example, for genetic analyses
in drug development pipelines. In this work, we consider an alternative
frequentist approach to colocalization testing that examines the proportionality
of genetic associations with each trait. The proportional colocalization
approach uses markedly different assumptions to Bayesian colocalization
tests, and therefore can provide valuable complementary evidence in
cases where Bayesian colocalization results are inconclusive or sensitive
to priors. We propose a novel conditional test of proportional colocalization,
\textsl{prop-coloc-cond}, that aims to account for the uncertainty
in variant selection, in order to recover accurate type I error control.
The test can be implemented straightforwardly, requiring only summary
data on genetic associations. Simulation evidence and an empirical
investigation into \textsl{GLP1R} gene expression demonstrates how
tests of proportional colocalization can offer important insights
in conjunction with Bayesian colocalization tests. 
\end{abstract}

\section{Introduction}

Genome-wide association studies (GWASs) have revealed that a large
number of genetic variants are robustly associated with a wide range
of diseases and disease traits. This has motivated the development
of post-GWAS methods that use genetic association data to investigate
causal mechanisms of potentially related traits. Colocalization analyses
specifically aim to identify whether or not two traits are affected
by the same or distinct causal variants in a single genetic region;
the two traits are said to \textsl{colocalize} if they share the same
causal variants. The statistical challenge of such analyses is to
appropriately account for issues that could otherwise lead to false
positives (e.g. falsely concluding that the traits colocalize when
they do not), such as sampling uncertainty in measured genetic associations,
and the correlation between genetic variants.

To date, most of the popular methods for colocalization analyses can
be placed into one of two categories: enumeration colocalization and
proportional colocalization. Enumeration colocalization methods \citep{Giambartolomei2014,Wallace2020}
evaluate the evidence for colocalization from a Bayesian perspective,
and return posterior probabilities of colocalization and other competing
hypotheses. The methods require investigators to specify a prior probability
for colocalization, which is typically set to limit the potential
for false positives. The key features of this approach are that the
analyses can be performed using only summary data that is typically
shared in GWASs, and that the methods conclude that there is colocalization
only if there is evidence to refute the prior of no colocalization.

In contrast, proportional colocalization approaches \citep{Plagnol2009,Wallace2013}
are frequentist methods that test a null hypothesis of proportional
colocalization, a particular type of colocalization where the association
of causal variants with one trait are proportional to their associations
with the other trait (Figure 1). Aside from testing for only a specific
type of colocalization, the proportional testing approach also differs
from the enumeration approach in that the null hypothesis is that
the two traits colocalize. Therefore, whereas the enumeration colocalization
approach effectively looks for evidence against the prior of no colocalization,
the proportional testing approach looks for evidence against the null
of proportional colocalization. Note that to credibly accept the null
hypothesis of proportional colocalization we require that the proportional
colocalization test has high power. 
\begin{center}
\includegraphics[width=16.5cm]{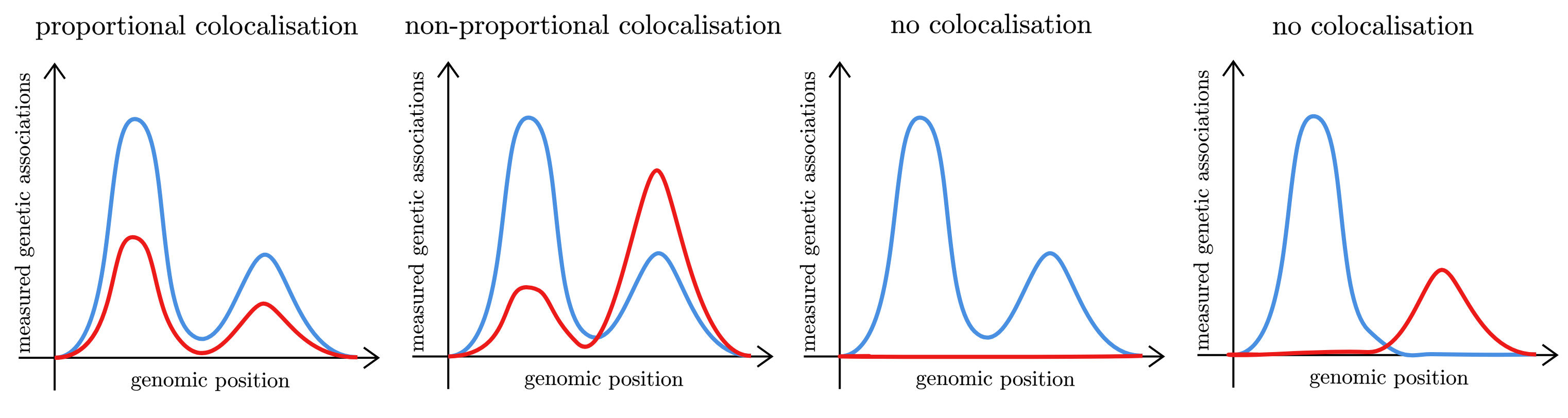}\\
\emph{\small{}Figure 1.}\emph{ }\emph{\footnotesize{}Measured genetic
associations in a single gene region in proportional, non-proportional,
and no colocalization settings. The red curve corresponds to trait
1, and the blue curve corresponds to trait 2. }{\footnotesize\par}
\par\end{center}

While the enumeration colocalization approach more directly evaluates
evidence for the general colocalization hypothesis, the proportional
colocalization hypothesis is at times of independent interest. The
proportionality of genetic associations with each trait would be consistent
with one trait mediating the other. Moreover, proportional colocalization
tests are closely related to overidentification tests \citep{Hansen1982}
and conditional F-tests \citep{Sanderson2016} which are routinely
used to indicate sources of possible model mis-specification in drug
target \textsl{cis}-Mendelian randomization analyses \citep{Burgess2023}. 

An important decision when performing a proportional colocalization
test concerns variant selection. The test effectively considers the
totality of evidence across all included variants, and therefore \textsl{aims}
to be sensitive to even slight departures from proportionality of
weak variant associations. In practice, this results in a tendency
to reject the proportional colocalization hypothesis when an included
variant is only weakly associated with one trait, even when the two
traits share another strongly associated causal variant. Enumeration
colocalization tests are less sensitive in this regard since they
aim to compare colocalization evidence of only strongly associated
variants.

In order to avoid sensitivity to weak variant associations, \citet{Wallace2013}
computes the proportional colocalization test with lead variants (the
variants most strongly associated with each trait), and highlights
the heavily inflated type I error rates that can result from ignoring
the uncertainty in variant selection. In this work, we aim to mitigate
this type I error problem through conditional inference techniques.
Specifically, we propose a novel test of proportional colocalization
based on lead variants that conditions on the selection event \citep{Fithian2017},
and hence accounts for the uncertainty in variant selection. 

More generally, we discuss how subtle differences in underlying models
of genetic variant-trait effects and definitions of colocalization
can result in differing evidence between and within proportional and
enumeration colocalization approaches. Since they work with fundamentally
different assumptions, combining the evidence provided by the two
approaches offers the potential to make more robust inferences than
if the evidence from only one approach is considered. We demonstrate
this practice by comparing colocalization evidence in an empirical
application involving gene expression in different tissues in the
\textsl{GLP1R} gene region. R code to apply our methods is available
at \href{https://github.com/ash-res/prop-coloc/}{github.com/ash-res/prop-coloc/}. 

\section{Methods}

\subsection{Proportional colocalization}

\subsubsection{Linear model and summary data}

Two traits are said to proportionately colocalize if: (i) they share
the same causal genetic variants in a single gene region; and (ii)
the trait associations with those causal variants are proportional
to each other. When there is only one causal variant for both traits,
genetic associations with two traits that colocalize are necessarily
proportional, and hence there is no distinction between colocalization
and proportional colocalization. But in the general case of multiple
causal variants, proportional colocalization is a specific type of
colocalization, as highlighted in the two leftmost plots in Figure
1. 

We are interested in testing the null hypothesis that two traits $X_{1}$
and $X_{2}$ proportionately colocalize. Let $Z=(Z_{1}\ldots Z_{J})^{\prime}$
denote a $J$-vector of genetic variants that are possibly causal
for $X_{1}$ and/or $X_{2}$. The genetic variants may be moderately
correlated, but the variance-covariance matrix of $Z$, $var(Z)$,
is assumed to be invertible. 

For each trait $k$, we consider the following linear model with homoscedastic
errors, 
\begin{equation}
X_{k}=\alpha_{k}+\gamma_{k}^{\prime}Z+V_{k}
\end{equation}
where $\alpha_{k}$ is an unknown constant, $\gamma_{k}$ is a $J$-vector
of unknown coefficients, and $V=(V_{1}\,\,V_{2})^{\prime}$ is a vector
of random errors which satisfies $E[V\vert Z]=0_{2\times1}$ and $E[VV^{\prime}\vert Z]=\Sigma_{V}$,
where $\Sigma_{V}$ is some unknown $2\times2$ positive definite
matrix. 

We assume access to summary data (beta coefficients and standard errors)
from univariable $X_{k}$ on $Z_{j}$ linear regressions that are
typically shared by GWASs. We also assume knowledge of a $J\times J$
genetic correlation matrix, and the correlation between traits $X_{1}$
and $X_{2}$. We consider a one-sample setting where genetic associations
with $X_{1}$ are measured from the same sample as genetic associations
with $X_{2}$, however it is straightforward to adapt the analysis
considered here for a two-sample setting in which genetic associations
with each trait are measured from non-overlapping samples. 

Using only univariable regression summary data, Proposition 1 of Supplementary
Information shows that we can construct \textsl{multivariable} estimates
$(\widehat{\gamma}_{1},\widehat{\gamma}_{2})$ of $(\gamma_{1},\gamma_{2})$
that are assumed to satisfy the normality assumption 
\begin{equation}
\begin{bmatrix}\widehat{\gamma}_{1}\\
\widehat{\gamma}_{2}
\end{bmatrix}\sim N\left(\begin{bmatrix}\gamma_{1}\\
\gamma_{2}
\end{bmatrix},\Sigma_{\gamma}\big/n:=\begin{bmatrix}\Sigma_{\gamma,11} & \Sigma_{\gamma,12}\\
\Sigma_{\gamma,12}^{\prime} & \Sigma_{\gamma,22}
\end{bmatrix}\big/n\right),
\end{equation}
where $\Sigma_{\gamma}=var(Z)^{-1}\otimes\Sigma_{V}$, $n$ denotes
the sample size used to measure genetic associations with traits,
and $\otimes$ denotes the Kronecker product. The normality assumption
$(2)$ is justified by standard large sample arguments (Proof of Proposition
1, Supplementary Information). The strategy of modelling joint effects
of multiple genetic variants using a multivariate normal distribution
has previously been explored in fine-mapping investigations; see,
for example, \citet{Verzilli2008,Yang2012,Newcombe2016}.

\subsubsection{Proportional colocalization test statistic}

The null hypothesis of proportional colocalization is $H_{0}:\gamma_{1}=\gamma_{2}\eta_{0}$
uniquely for some $\eta_{0}\not=0$. Following \citet{Wallace2013},
the parameter $\eta_{0}$ is called the proportionality constant.
To test $H_{0}$, we assume that both $\gamma_{1}$ and $\gamma_{2}$
are non-zero vectors. Although in practice most variants in $Z$ may
not be causal for either trait, it is assumed that there is at least
one (not necessarily the same) causal variant in $Z$ for $X_{1}$
and $X_{2}$. Our aim is to use the normality assumption $(2)$ to
construct a test of the proportional colocalization null hypothesis
$H_{0}$, against the general alternative $H_{1}:\gamma_{1}\not=\gamma_{2}\eta$
for any $\eta$. 

An estimating function for the proportionality constant is $\widehat{g}(\eta)=\widehat{\gamma}_{1}-\widehat{\gamma}_{2}\eta$
since $\widehat{g}(\eta)$ has mean zero uniquely at $\eta_{0}$ under
$H_{0}$. Note that because we consider $J$ variants, there are $J$
estimating equations for the 1 unknown $\eta$; there is \textsl{no}
value of $\eta$ that can simultaneously satisfy all $J$ estimating
equations $\widehat{g}(\eta)=0$. Instead, the generalized method
of moments (GMM; \citealp{Hansen1982}) minimizes a normalized criterion
of estimating equations. Specifically, the continuously updating GMM
\citep{Hansen1996} criterion is given by $\widehat{Q}(\eta)=\widehat{g}(\eta)^{\prime}\Omega(\eta)^{-1}\widehat{g}(\eta)$,
where $\Omega(\eta)=\Sigma_{\gamma,11}-\Sigma_{\gamma,12}\eta-\Sigma_{\gamma,12}^{\prime}\eta+\Sigma_{\gamma,22}\eta^{2}$
is the large sample variance of $\widehat{g}(\eta)$. The corresponding
GMM estimator of the proportionality constant is $\widehat{\eta}=\arg\min_{\eta}\widehat{Q}(\eta)$. 

Under some GMM regularity conditions, it can be shown that in large
samples $n\widehat{Q}(\widehat{\eta})$ has a $\chi^{2}$ distribution
with $J-1$ degrees of freedom (Proposition 2, Supplementary Information).
Thus, a $\nu$-level proportional colocalization test rejects the
null hypothesis $H_{0}$ if $\min_{\eta}n\widehat{Q}(\eta)>q_{J-1}(\nu)$,
where $q_{J-1}(\nu)$ is the $(1-\nu)$-th quantile $q_{J-1}(\nu)$
of the $\chi^{2}$ distribution with $J-1$ degrees of freedom. 

\subsubsection{Practical issues for variant selection}

There are at least two important considerations when choosing which
genetic variants should be included in the proportional colocalization
test. First, the proportional colocalization hypothesis $H_{0}:\gamma_{1}=\gamma_{2}\eta_{0}$
requires that proportionality of trait associations hold for every
variant included in the test. As such, it considers the totality of
evidence across all included variants. The inclusion of many irrelevant
variants that are not causal for either trait is shown in Section
3.1.1 to harm the finite-sample size properties of the proportional
colocalization test. 

At the same time, selecting variants based on their measured associations
with either trait can also lead to inflated type I error rates if
the sampling uncertainty in variant selection is not accounted for.
When considering only the top 10 associated variants with each trait,
we find that the type I error rate of the proportional colocalization
test can be over 50\% for a 5\% level test (Supplementary Figures
S1 and S2). Although some variant selection is usually necessary,
it is seemingly difficult to make good inferences in finite samples
without adjusting for variant selection. 

Second, since the proportional colocalization test statistic is a
function of the inverse of the variance-covariance matrix of genetic
variants, the test can be numerically unstable when very highly correlated
variants are included. In general, there are no obvious guidelines
on how high correlations between selected variants should be: on the
one hand, allowing variant correlations up to only, say, a $R^{2}\leq0.4$
correlation threshold runs a higher risk that we are omitting true
causal variants from the analysis. However, allowing highly correlated
variants up to a $R^{2}\leq0.9$ threshold may be problematic, especially
if the available genetic correlation estimates are not representative
of the sample used to compute genetic associations with traits. For
example, this may be due to excessive uncertainty in estimating genetic
correlations, or if genetic correlation estimates are based on individuals
of different ethnicity to those represented in measured genetic variant--trait
associations. 

\subsection{Conditional proportional colocalization test}

\subsubsection{Naive tests}

We now revisit the strategy of \citet{Wallace2013}, which tests the
proportionality of trait associations using only two lead variants.
The lead variants are selected based on their \textsl{multivariable}
associations with traits (i.e. based on $X_{k}$ on $Z=(Z_{1},\ldots,Z_{J})^{\prime}$
linear regressions rather than univariable $X_{k}$ on $Z_{j}$ linear
regressions), and hence the measured association of any given variant
is adjusted for all other variants. To do this, we start by using
the normality assumption $(2)$ to construct simple $t$-statistics
to judge the relevance of a given variant for each trait. 

For each trait $k$, let $D_{\gamma,k}$ be the $J\times J$ diagonal
matrix such that its $(j,j)$-th element is given by $(\Sigma_{\gamma,kk})_{jj}^{-1/2}$
where $(\Sigma_{\gamma,kk})_{jj}$ is the $(j,j)$-th element of $\Sigma_{\gamma,kk}$.
Next, let $D_{\gamma}$ be the $2J\times2J$ diagonal matrix such
that its top-left $J\times J$ block is equal to $D_{\gamma,1}$,
and its bottom-right $J\times J$ block is equal to $D_{\gamma,2}$.
Then, $\widehat{T}=(\widehat{T}_{1}^{\prime}\,\,\widehat{T}_{2}^{\prime})^{\prime}=D_{\gamma}\sqrt{n}(\widehat{\gamma}_{1}^{\prime},\widehat{\gamma}_{2}^{\prime})^{\prime}$
denotes a $2J$-vector of $t$-statistics, where $\widehat{T}_{k}$
is the $J$-vector of $t$-statistics corresponding to trait $k$,
$(k=1,2)$. 

We first select the $j^{\star}$-th variant that is most strongly
associated with trait 1, so that $\vert\widehat{T}_{1j^{\star}}\vert\geq\vert\widehat{T}_{1j}\vert$
for all variants $j=1,\ldots,J$, where, for example, $\widehat{T}_{1j}$
is the $t$-statistic of the association of the $j$-th variant with
trait 1. Then, from the remaining variants not including the $j^{\star}$-th
variant, we select the $j^{\star\star}$-th variant that is most strongly
associated with trait 2, so that $\vert\widehat{T}_{2j^{\star\star}}\vert\geq\vert\widehat{T}_{2j}\vert$
for all variants $j=\{1,\ldots,J\}\backslash\{j^{\star}\}$, $j^{\star\star}\not=j^{\star}$.
We note that this procedure could induce non-commutativity; if both
traits share the same strongest variant, the second strongest variant
for trait 2 may not be the same as the second strongest variant for
trait 1. 

To compute the proportional colocalization test statistic based on
only these two selected variants, let $I_{\star}$ denote the $2\times J$
matrix with its $(1,j^{\star})$-th and $(2,j^{\star\star})$-th elements
equal to $1$, and all other elements equal to $0$. Then, as in Section
2.1.2, we can define the GMM quantities $\widehat{g}_{\star}(\eta)=I_{\star}\widehat{g}(\eta)$,
$\Omega_{\star}(\eta)=I_{\star}\Omega(\eta)I_{\star}^{\prime}$, and
$\widehat{Q}_{\star}(\eta)=\widehat{g}_{\star}(\eta)^{\prime}\Omega_{\star}(\eta)^{-1}\widehat{g}_{\star}(\eta)$.
An estimator of the proportionality constant is $\widehat{\eta}_{\star}=\arg\min_{\eta}\widehat{Q}_{\star}(\eta)$,
and a \textsl{naive} $\nu$-level proportional colocalization test
that ignores the uncertainty in the variant selection step simply
compares the statistic $n\widehat{Q}_{\star}(\widehat{\eta})$ against
the $(1-v)$-th quantile $q_{1}(\nu)$ of the $\chi^{2}$ distribution
with 1 degree of freedom. We refer to this test as \textsl{prop-coloc-naive}. 

\subsubsection{Adjusting for variant selection with conditional critical values}

If there was no uncertainty in variant selection, the limiting $\chi^{2}$
distribution of the proportional colocalization test statistic $n\widehat{Q}_{\star}(\widehat{\eta})$
is established by noting that $n\widehat{Q}_{\star}(\widehat{\eta})$
is closely approximated by $\big(\Omega_{\star}^{-1/2}\sqrt{n}\widehat{g}_{\star}(\eta_{0})\big)^{\prime}M_{\star}\big(\Omega_{\star}^{-1/2}\sqrt{n}\widehat{g}_{\star}(\eta_{0})\big)$
in large samples, where $\Omega_{\star}=\Omega_{\star}(\eta_{0})$,
$G_{\star}=-I_{\star}\gamma_{2}$, and $M_{\star}=I_{2\times2}-\Omega_{\star}^{-1/2}G_{\star}(G_{\star}^{\prime}\Omega_{\star}^{-1}G)^{-1}G_{\star}^{\prime}\Omega_{\star}^{-1/2}$
is an idempotent matrix of rank $1$ (Proof of Proposition 2, Supplementary
Information). Under $H_{0}$ and the normality assumption $(2)$,
$\Omega_{\star}^{-1/2}\sqrt{n}\widehat{g}_{\star}(\eta_{0})$ is distributed
as a $N(0_{2\times1},I_{2\times2})$ random variable, so that in large
samples, the statistic $\big(\Omega_{\star}^{-1/2}\sqrt{n}\widehat{g}_{\star}(\eta_{0})\big)^{\prime}M_{\star}\big(\Omega_{\star}^{-1/2}\sqrt{n}\widehat{g}_{\star}(\eta_{0})\big)$
converges to a $\chi^{2}$ random variable with $1$ degree of freedom. 

To consider how the uncertainty in variant selection may affect the
distribution of the proportional colocalization test statistic, we
model the dependence of the statistic $\Omega_{\star}^{-1/2}\sqrt{n}\widehat{g}_{\star}(\eta_{0})$
on the vector of $t$-test statistics $\widehat{T}$ which determines
variant selection. Specifically, we assume that under $H_{0}:\gamma_{2}=\gamma_{1}\eta_{0}$,
the normalized statistics $\Omega_{\star}^{-1/2}\sqrt{n}\widehat{g}_{\star}(\eta_{0})$
and $\widehat{T}$ are jointly normal with covariance $cov(\Omega_{\star}^{-1/2}\sqrt{n}\widehat{g}_{\star}(\eta_{0}),\widehat{T})=C_{\star}$,
where $C_{\star}=\Omega_{\star}^{-1/2}I_{\star}C\Sigma_{\gamma}D_{\gamma}$
and $C=[\Sigma_{\gamma,11}-\Sigma_{\gamma,12}^{\prime}\eta_{0},\Sigma_{\gamma,12}-\Sigma_{\gamma,22}\eta_{0}]$.
Since $\Omega_{\star}^{-1/2}\sqrt{n}\widehat{g}_{\star}(\eta_{0})$
has mean zero under $H_{0}$, the joint distribution of $\Omega_{\star}^{-1/2}\sqrt{n}\widehat{g}_{\star}(\eta_{0})$
and $\widehat{T}$ then depends on only $C_{\star}$ and the mean
of $\widehat{T}$, which we denote by $T$. 

Since $T$ is unknown, we choose to condition on a sufficient statistic
${\cal L}$ for $T$, where ${\cal L}=\widehat{T}-C_{\star}^{\prime}\Omega_{\star}^{-1/2}\sqrt{n}\widehat{g}_{\star}(\eta_{0})$.
Note that this statistic is uncorrelated with $\Omega_{\star}^{-1/2}\sqrt{n}\widehat{g}_{\star}(\eta_{0})$,
and that under $H_{0}$,
\begin{equation}
\begin{bmatrix}\Omega_{\star}^{-1/2}\sqrt{n}\widehat{g}_{\star}(\eta_{0})\\
\widehat{T}
\end{bmatrix}\Big|({\cal L}=\ell)\sim\begin{bmatrix}{\cal K}\\
\ell+C_{\star}^{\prime}{\cal K}
\end{bmatrix},\,\,\,\text{where}\,\,{\cal K}\sim N(0_{2\times1},I_{2\times2}).
\end{equation}
Equation $(3)$ provides the basis to approximate the conditional
distribution of the proportional colocalization test statistic, and
hence to adjust inferences for the uncertainty in variant selection.
In particular, we condition on the variant selection event \citep{Fithian2017},
which is given by ${\cal S}=\cap_{j\in[J]}\{\vert\widehat{T}_{1j^{\star}}\vert\geq\vert\widehat{T}_{1j}\vert\}\cap_{j\in[J]\backslash\{j^{\star}\}}\{\vert\widehat{T}_{2j^{\star\star}}\vert\geq\vert\widehat{T}_{2j}\vert\}$.
Therefore, in contrast with the naive test described in Section 2.2.1,
the conditional test does \textsl{not} compare the statistic $n\widehat{Q}_{\star}(\widehat{\eta})$
against $q_{1}(\nu)$, but rather a critical value $w_{\nu}^{\star}$
such that $P_{H_{0}}(n\widehat{Q}_{\star}(\widehat{\eta})\leq w_{\nu}^{\star}\vert{\cal S},{\cal L}=\ell)=1-\nu$
under $H_{0}$. If $n\widehat{Q}_{\star}(\widehat{\eta})>w_{\nu}^{\star}$,
then we reject the null hypothesis of proportional colocalization.
We refer to this method as \textsl{prop-coloc-cond}. 

In order to implement the prop-coloc-cond test in practice, we can
use $(3)$ to approximate the conditional distribution $P_{H_{0}}(n\widehat{Q}_{\star}(\widehat{\eta})\leq w\vert{\cal S},{\cal L}=\ell)$,
which can then be used to find the critical value $w_{\nu}^{\star}$.
Let $\ell_{{\cal K}}=\widehat{T}\vert({\cal L}=\ell)$, and as with
the $2J$-vector of $t$-statistics $\widehat{T}=(\widehat{T}_{1}^{\prime},\widehat{T}_{2}^{\prime})^{\prime}$,
partition $\ell_{{\cal K}}=\ell+C_{\star}^{\prime}{\cal K}$ into
$\ell_{{\cal K}}=(\ell_{{\cal K},1}^{\prime},\ell_{{\cal K},2}^{\prime})^{\prime}$
where $\ell_{{\cal K},k}=(\ell_{{\cal K},k1},\ldots,\ell_{{\cal K},kJ})^{\prime}$
for each trait $k=1,2$. Then, the conditional distribution
\[
P_{H_{0}}(n\widehat{Q}_{\star}(\widehat{\eta})\leq w\vert{\cal S},{\cal L}=\ell)=\frac{P_{H_{0}}(\{n\widehat{Q}_{\star}(\widehat{\eta})\leq w\}\cap{\cal S}\vert{\cal L}=\ell)}{P_{H_{0}}({\cal S}\vert{\cal L}=\ell)}
\]
is approximately
\begin{equation}
\frac{P\Big(\{{\cal K}^{\prime}M_{\star}{\cal K}\leq w\}\cap_{j\in[J]}\{\vert\ell_{{\cal K},1j^{\star}}\vert\geq\vert\ell_{{\cal K},1j}\vert\}\cap_{j\in[J]\backslash\{j^{\star}\}}\{\vert\ell_{{\cal K},2j^{\star\star}}\vert\geq\vert\ell_{{\cal K},2j}\vert\}\Big)}{P\Big(\cap_{j\in[J]}\{\vert\ell_{{\cal K},1j^{\star}}\vert\geq\vert\ell_{{\cal K},1j}\vert\}\cap_{j\in[J]\backslash\{j^{\star}\}}\{\vert\ell_{{\cal K},2j^{\star\star}}\vert\geq\vert\ell_{{\cal K},2j}\vert\}\Big)}.
\end{equation}
We can estimate $(4)$ by taking many draws of ${\cal K}\sim N(0_{2\times1},I_{2\times2})$,
and in practice we evaluate the conditional distribution at the fitted
value $\ell=\widehat{T}-C_{\star}^{\prime}\Omega^{-1/2}\sqrt{n}\widehat{g}_{\star}(\widehat{\eta})$.
In a sparse effects setting where it is obvious which are the lead
variants, there should be little uncertainty in variant selection
and hence the conditional critical value $w_{\nu}^{\star}$ should
be very close to the standard critical value $q_{1}(\nu)$ based on
the $(1-\nu)$-th quantile of the $\chi^{2}$ distribution with 1
degree of freedom. However, in more general cases, the conditional
critical value will adjust to accommodate the sampling uncertainty
in determining the lead variants.

\subsubsection{Inference on the proportionality constant}

Since the proportionality of genetic variant--trait associations
$\gamma_{1}=\gamma_{2}\eta_{0}$ trivially holds when $\gamma_{2}\not=0$
and $\eta_{0}=0$, the proportional colocalization test has no power
to detect no colocalization when there is a causal variant for trait
2 but no causal variant for trait $1$. In order to rule out colocalization
in this case, it would be helpful to also test the hypothesis $H_{0,\eta}:\eta_{0}=0$
against the general alternative $H_{1,\eta}:\eta_{0}\not=0$. For
this, we consider using a Lagrange multiplier (LM; for example, \citealp{Smith1997})
test that involves no estimation of $\eta_{0}$. 

Let $\widehat{G}_{\star}=-I_{\star}\widehat{\gamma}_{2}$. Under proportional
colocalization $\gamma_{2}=\gamma_{1}\eta_{0}$ and $H_{0,\eta}:\eta_{0}=0$,
the LM test statistic is given by 
\[
LM=\frac{n(\widehat{G}_{\star}^{\prime}\Omega_{\star}(0)^{-1}\widehat{g}_{\star}(0))^{2}}{\widehat{G}_{\star}^{\prime}\Omega_{\star}(0)^{-1}\widehat{G}_{\star}},
\]
and, taking variant selection as given, the statistic converges to
a $\chi^{2}$ random variable with 1 degree of freedom as $n\to\infty$
(Proposition 3, Supplementary Information). Therefore, a $\nu$-level
test of $H_{0,\eta}$ compares the statistic $LM$ with $q_{1}(\nu)$,
the $(1-\nu)$-th quantile of the $\chi^{2}$ distribution with 1
degree of freedom. If $LM>q_{1}(\nu)$, then we reject $H_{0,\eta}$
and conclude evidence for a non-zero proportionality constant, and
hence the presence of a causal variant for trait 1. Although this
inference procedure does not explicitly account for variant selection,
this appears to have a negligible impact on the size performance of
the test. Supplementary Figures S3 and S4 verify that this test accurately
controls type I error rates, and has power to detect a non-zero proportionality
constant in small and large samples. 

\subsection{Comparisons with Bayesian colocalization tests}

Before presenting a more comprehensive simulation study, we briefly
discuss some toy numerical examples of scenarios where we may expect
differences between the results of proportional and enumeration colocalization
approaches. 

We note the results of two enumeration colocalization tests. First,
the ``coloc'' \citep{Giambartolomei2014,Wallace2020} test is based
on an assumption that there is at most a single causal variant for
each trait, and provides posterior probabilities of colocalization
and other competing hypotheses. Specifically, under ``H0'' there
is no causal variant for either trait, under ``H1'' there is a causal
variant only for trait 1, under ``H2'' there is a causal variant
only for trait 2, and under ``H3'' each trait has a different causal
variant. The hypothesis of interest is \textquotedblleft H4\textquotedblright ,
which is that the two traits have a shared causal variant. Second,
we also consider a recently proposed enumeration colocalization method
that combines coloc with a Sum of Single Effects (``coloc-SuSiE'';
\citealp{Wallace2021}) regression framework; the method has built-in
variant selection, and can report the evidence for colocalization
at multiple variant loci. 

We compute coloc and coloc-SuSiE with default priors using the \texttt{coloc}
R package. For coloc, we take a posterior probability of H4 lower
than 0.5 to conclude evidence of no colocalization, and for coloc-SuSiE
we take a posterior probability of H4 lower than 0.5 at all variant
loci to conclude evidence of no colocalization. 

These enumeration colocalization approaches are compared with two
versions of proportional colocalization tests: ``\emph{prop-coloc-full}''
computes the proportional colocalization test discussed in Section
2.1.2 comparing the proportionality of all variant associations included
in the test, and ``\emph{prop-coloc-cond}'' computes the conditional
test of proportional colocalization of lead variant associations discussed
in Section 2.2.2. The nominal size of the proportional colocalization
tests was set at 0.05. 

We generated data based on the linear model $(1)$ in Section 2.1.1
for 40 uncorrelated genetic variants, with their true effects $\gamma_{1}$
and $\gamma_{2}$ on the two traits plotted in Figure 2. The sample
size was set to $n=1000$. 

In Model 1 of Figure 2, there is only a single causal variant for
trait 2, which is also causal for trait 1, and all other variants
have weak effects only on trait 1. Since the proportionality of variant--trait
associations does not hold over all variants or any two lead variants,
prop-coloc-full and prop-coloc-cond tend to reject the proportional
colocalization hypothesis. There is a shared causal variant that is
strongly associated with trait 2; in this case coloc tends to favour
the colocalization hypothesis, whereas coloc-SuSiE does not. 
\begin{center}
\includegraphics[width=16cm]{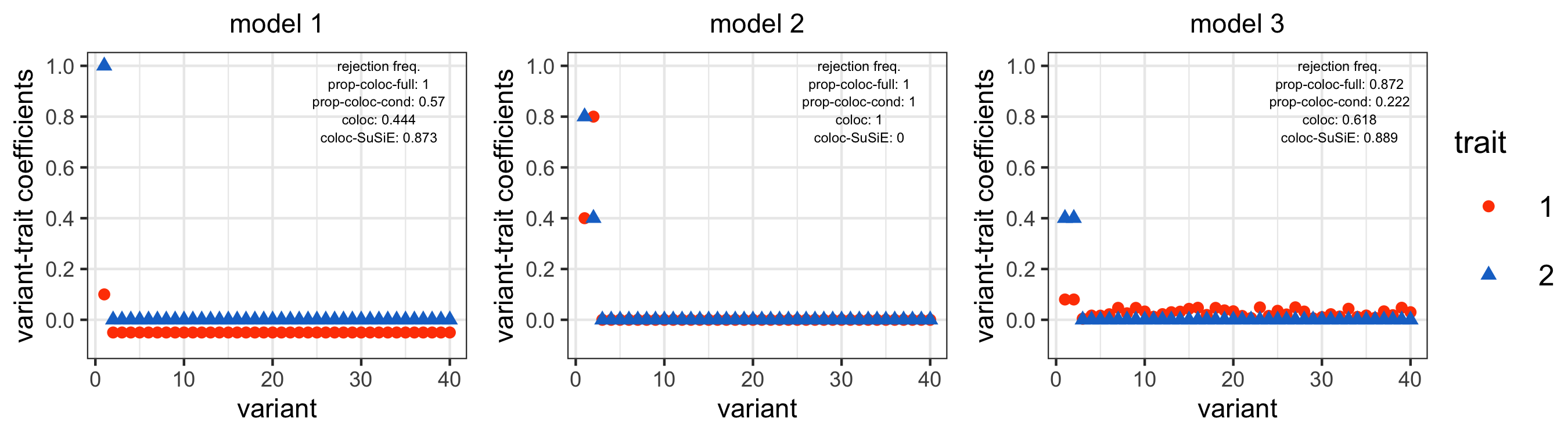}\\
\emph{\small{}Figure 2.}\emph{ }\emph{\footnotesize{}Colocalization
test results under three different model designs involving 40 uncorrelated
variants. These are merely illustrative examples to highlight potential
scenarios where different colocalization approaches can conclude in
differing evidence. The nominal size of the proportional colocalization
tests was set to 0.05, and a rejection of the colocalization hypothesis
using coloc and coloc-SuSiE was defined as the posterior probability
of H4 being lower than 0.5.}{\footnotesize\par}
\par\end{center}

Model 2 is a setting of non-proportional colocalization with two causal
variants that are strongly associated with both traits. In this case,
only coloc-SuSiE detects colocalization. Finally, Model 3 is a setting
where two variants are strongly associated with trait 2 and are weakly
associated with trait 1. All other variants have even weaker random
effects on only trait 1. These random effects tend to force prop-coloc-full
into rejecting the proportional colocalization hypothesis, whereas
prop-coloc-cond does not since the proportionality of trait associations
holds between the first two lead variants. Since all variant effects
on trait 1 are quite small, both enumeration colocalization tests
tend to favour a hypothesis of a causal variant for trait 2 only. 

\subsection{Simulation study}

We investigate the finite-sample performance of three versions of
proportional colocalization tests: ``prop-coloc-full'', ``prop-coloc-naive'',
and ``prop-coloc-cond''. For prop-coloc-full, the full set of genetic
variants is included in the proportional colocalization test as in
Section 2.1.2. Both prop-coloc-naive and prop-coloc-cond use the proportional
colocalization test statistic based on the two strongest variant associations
with the two traits, but prop-coloc-cond uses a critical value that
attempts to account for the uncertainty in variant selection as described
in Section 2.2.2, whereas the prop-coloc-naive uses the standard critical
value based on the $\chi^{2}$ distribution as described in Section
2.2.1. 

To aid a straightforward comparison, we primarily consider the setting
of a single causal variant for each trait, so that the proportional
colocalization hypothesis with a non-zero proportionality constant
is exactly the same as the general colocalization hypothesis studied
by enumeration colocalization tests. As in Section 2.3, we compute
the ``coloc'' \citep{Giambartolomei2014} and ``coloc-SuSiE''
\citep{Wallace2021} tests with default priors using the \texttt{coloc}
R package, and we take a posterior probability of colocalization lower
than 0.5 (at all variant loci for coloc-SuSiE) to conclude evidence
of no colocalization. For the three proportional colocalization tests,
the nominal size of the tests was set at $\nu=0.05$. The setting
of multiple causal variants is discussed in Section 3.1.3 below. 

The goal of the simulation study is to investigate the finite-sample
size and power properties of proportional colocalization tests, with
a particular emphasis on verifying the type I error control of the
prop-coloc-cond test. 

\subsubsection{Simulation design}

We generated genetic association data on $J=40$ genetic variants
such that there is only a single causal variant for each trait. The
variants $Z=(Z_{1},\ldots Z_{40})^{\prime}$ were generated from the
joint normal distribution $Z\sim N(0_{40\times1},\rho)$, where $\rho$
was set to be an invertible correlation matrix where the diagonal
elements were set equal to 1 so that all variants have equal variance.
The off-diagonal elements were set to be the off-diagonal elements
of the matrix $aa^{\prime}$ for $a=(a_{1},\ldots,a_{40})^{\prime}$,
where $a_{j}\sim U[0,\sqrt{\rho_{0}}]$ for $j=1,\ldots,40$, and
where $\rho_{0}<1$ is a positive constant. We set $\rho_{0}=0.8$
so that the correlation between any two distinct variants in $Z$
was at most $R^{2}=0.64$, with the exception of causal variants.
The correlation between the causal variant for trait $1$ and the
causal variant for trait $2$ was set equal to $\xi$. When $\xi=0$
the causal variants for each trait are uncorrelated, and $\xi=1$
represents the case where the traits share the same causal variants
(i.e. the traits colocalize). 

We generated traits from the linear model described in Section 2.1.1,
where the intercepts $(\alpha_{1},\alpha_{2})$ were set to be zero,
and the errors of the linear model were generated as $V=(V_{1},V_{2})\sim N(0_{2\times1},\Sigma_{V})$
where $\Sigma_{V}$ is the correlation matrix such that $cov(V_{1},V_{2})=0.3$.
The variant effects $\gamma_{1}$ on trait $1$ were set as $0.5$
for the causal variant, and $0$ for non-causal variants. The variant
effects $\gamma_{2}$ on trait $2$ were set as $0.5\eta_{0}$ for
the causal variant, and $0$ for non-causal variants. 

From an $n$-sized sample on $X_{1}$, $X_{2}$, and $Z$, we generated
univariable summary data on genetic associations. For computing the
coloc and proportional colocalization test results, we used the estimated
coefficients and corresponding standard errors from $X_{k}$ on $Z_{j}$
linear regressions for each $k=1,2$, and $j=1,\ldots,40$, as well
as the sample correlation matrix of $Z$. In addition, proportional
colocalization tests used the sample correlation between $X_{1}$
and $X_{2}$, and knowledge of the sample size $n$. 

\subsection{Empirical application}

We consider a colocalization analysis of gene expression in different
tissues in the \textsl{GLP1R} gene region. Such analyses are of potentially
of interest when investigating causal mechanisms by which GLP1R agonists
affect disease risk \citep{Daghlas2021,Patel2023}. 

\subsubsection{Genetic variant associations with \textsl{GLP1R} gene expression}

Estimated genetic associations with gene expression based on $n=838$
participants of mostly European ancestry were taken from the Genotype-Tissue
Expression (GTEx) project version 8 \citep{GTExConsortium2020}. Estimated
genetic variant correlations based on 367,703 unrelated participants
were taken from the UK Biobank \citep{Astle2016}. We considered the
genetic region \textpm{} 100kbp of the \textsl{GLP1R} gene (chr6:39,016,557-39,059,079
in GRCh37/hg19) for which 851 variant associations with \textsl{GLP1R}
expression were available. 

Our analysis studies \textsl{GLP1R} expression in 10 tissues that
were used to fit a multivariable model for coronary artery disease
risk in \citet{Patel2023}. The tissues are thyroid, testis, stomach,
pancreas, nerve, lung, left ventricle (heart), atrial appendage (heart),
hypothalamus (brain), and caudate (brain). Our colocalization analyses
involve considering as traits \textsl{GLP1R} expression in any 2 of
these 10 tissues in turn. 

For computing proportional colocalization tests, we pruned variants
up to a correlation threshold of $R^{2}\leq0.6$, and after this we
considered only the top 10 associated variants with each trait (measured
by marginal p-values). Such selection on p-values may harm the size
properties of the prop-coloc-full test, but appears to have a relatively
low impact on the prop-coloc-cond test (Supplementary Figures S1 and
S2). The trait with the strongest variant association (lowest p-value)
was selected as trait 2. 

We present the results of ``prop-coloc-cond-LM'' which combines
the results of the prop-coloc-cond test with the results of a Lagrange
multiplier (LM) test for a non-zero proportionality constant, as discussed
in Section 2.2.3. Specifically, the p-value of the prop-coloc-cond
test is presented only for trait comparisons where the LM test concludes
in strong evidence (at the 95\% confidence level) for a non-zero proportionality
constant. Otherwise, the proportional colocalization hypothesis is
rejected regardless of the p-value of the prop-coloc-cond test because
the LM test suggests that there is insufficient evidence for a causal
variant for trait 1. The results of ``prop-coloc-full-LM'' are defined
analogously from combining the prop-coloc-full and LM test results. 

The proportional colocalization tests require an input of trait correlations;
the trait correlations were set to 0, with the exception of the two
brain tissues and the two heart tissues which were set at 0.5. The
results were not too sensitive to this choice. As in the simulation
study, the default priors were selected for the coloc and coloc-SuSiE
tests. 

\section{Results}

\subsection{Simulation study results}

\subsubsection{Single causal variant}

Figure 3 presents the type I error and power results of the colocalization
tests for the single causal variant case, varying with the sample
size $n$ (from the top row to the bottom row), the proportionality
constant $\eta_{0}$ (from the left column to the right column), and
the correlation between the causal variants of trait $1$ and trait
$2$ (on the x-axis of each plot). 

As expected, the naive test that does not account for variant selection
uncertainty can have very poor size properties even in large sample
settings; for example, the test has a type I error rate of 0.26 for
the case of $\eta_{0}=1$ and $n=10000$. In contrast, the conditional
proportional colocalization test is able to control type I error rates
near the nominal 0.05 level, even though it is slightly over-sized
in some cases. The proportional colocalization test using the full
set of $40$ variants has inflated type I error rates compared with
the conditional test in small sample settings, with its type I error
rates consistently over 0.1 when $n=500$ for all values of the proportionality
constant $\eta_{0}$ considered. 

Similar to the simulation evidence in \citet{Wallace2013}, the correlation
between causal variants is shown to affect the power properties of
proportional colocalization tests, as well as the type I error rate
of coloc tests in finite samples. Intuitively, as two distinct causal
variants become more correlated, it becomes harder to separate the
two genetic signals, leading to lower rejection rates of the proportional
colocalization test. For the case of $\eta_{0}=0.5$ and $n=1000$,
the power of prop-coloc-cond falls from around 0.9 when $\xi=0.7$
to around 0.4 when $\xi=0.9$ when $n=1000$ and $\eta_{0}=0.5$.
Similarly, coloc is more likely to falsely conclude that there is
colocalization for highly correlated distinct causal variants in small
samples; for $\eta_{0}=0.5$ and $n=500$, the coloc type I error
rate is less than 0.05 when $\xi=0.4$ but around 0.6 when $\xi=0.8$. 

The results in Figure 3 also suggest specific situations where we
could expect differences in the evidence provided by coloc and proportional
colocalization tests. When the proportionality constant $\eta_{0}$
is close to $0$, this represents a setting where the genetic signal
for trait $1$ is relatively weak. The tendency of coloc in this case
is to favour the hypothesis of a causal variant for only trait 2 in
small samples, whereas the proportional colocalization test tends
to retain the null hypothesis of colocalization. This is useful from
the perspective that the proportional testing approach is less likely
to erroneously reject the null of colocalization when the signal for
trait 1 is weak. However, since the proportional colocalization hypothesis
$H_{0}:\gamma_{2}=\gamma_{1}\eta_{0}$ is trivially satisfied when
$\eta_{0}=0$, proportional colocalization tests have no power to
reject colocalization when there is no genetic signal for trait 1.
Hence, proportional colocalization tests should be used only when
there is strong evidence for a genetic signal for both traits in the
gene region.
\begin{center}
\includegraphics[width=16.5cm]{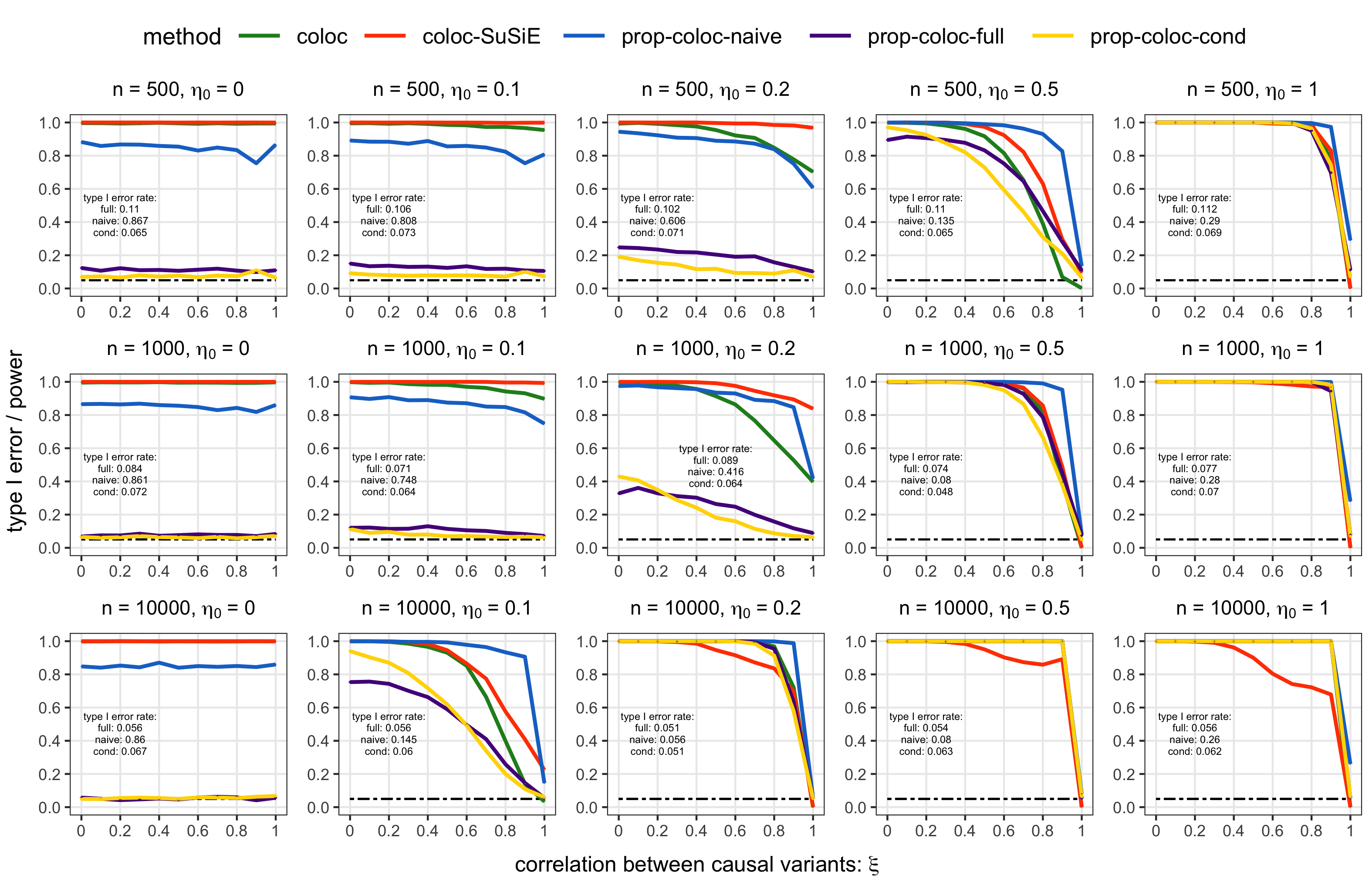}\\
\emph{\small{}Figure 3.}\emph{ }\emph{\footnotesize{}The rejection
frequencies of colocalization tests, with a single causal variant
for traits 1 and 2. The traits colocalize only where the causal variants
are perfectly correlated $(\xi=1)$. The nominal size of proportional
colocalization tests was set at 0.05. For coloc and coloc-SuSiE the
rejection rates of the colocalization hypothesis, $P(H4<0.5)$, are
plotted.}{\footnotesize\par}
\par\end{center}

\subsubsection{Mis-measured genetic correlations}

An important problem in practice is that investigators may not always
have access to genetic correlation estimates from the same sample
used to compute genetic assocations with traits. In such situations,
genetic correlations from a non-overlapping reference sample may be
used, but the performance of the proportional colocalization test
with many variants may be sensitive to errors in genetic correlation
estimates. One strategy to ensure the test is not too vulnerable to
mis-measured genetic correlations is to consider only moderately correlated
variants through pruning. Pruning is a stepwise procedure that finds
a subset of variants that are as strongly associated with traits as
possible, up to the constraint that the mutual correlation between
any two variants in that subset does not exceed some user specified
threshold \citep{Dudbridge2016}. 

To investigate the impact of mis-measured genetic correlations on
proportional colocalization testing, we consider the same simulation
design described in Section 2.4.1, but now we use noisy genetic correlation
estimates from a reference sample when computing proportional colocaliztion
tests. In particular, instead of using the sample genetic correlation
matrix, we use a random Wishart matrix centered at the true genetic
correlation matrix with degrees of freedom $\lambda$. We also compute
the prop-coloc-full test at different pruning thresholds based on
the $R^{2}$ correlation between variants. 
\begin{center}
\includegraphics[width=16.5cm]{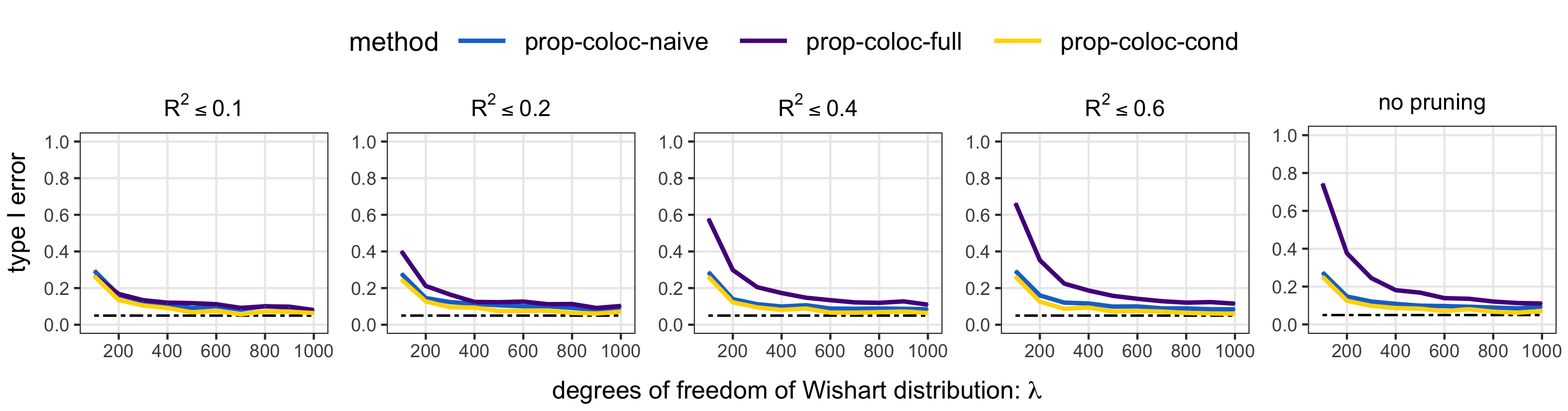}\\
\emph{\small{}Figure 4.}\emph{ }\emph{\footnotesize{}Type I error
rates of proportional colocalization tests varying with estimation
errors in genetic correlations. Lower degrees of freedom $\lambda$
correspond to greater errors in correlation estimates. The nominal
size of the tests was set at $0.05$. }{\footnotesize\par}
\par\end{center}

Figure 4 plots the type I error rates of proportional colocalization
tests for the case where the sample size is $n=1000$ and the proportionality
constant is $\eta_{0}=0.5$. For the case of no pruning, where the
correlation of any two variants may be up to $R^{2}=0.8^{2}$ as described
in Section 2.4.1, the type I error rates of prop-coloc-full are considerably
higher than prop-coloc-cond. The type I error problem becomes more
severe for low enough degrees of freedom $\lambda$, since lower values
of $\lambda$ correspond to larger errors in genetic correlation estimates. 

The performance of the prop-coloc-full test based on different pruning
thresholds is shown in the first 4 panels of Figure 4; the prop-coloc-cond
and prop-coloc-naive results are also shown for comparison. Aggressive
pruning seems to alleviate the inflated type I error rates of the
prop-coloc-full test, with a pruning threshold of $R^{2}\leq0.1$
leading to similar size performance as prop-coloc-cond. 
\begin{center}
\includegraphics[width=16.5cm]{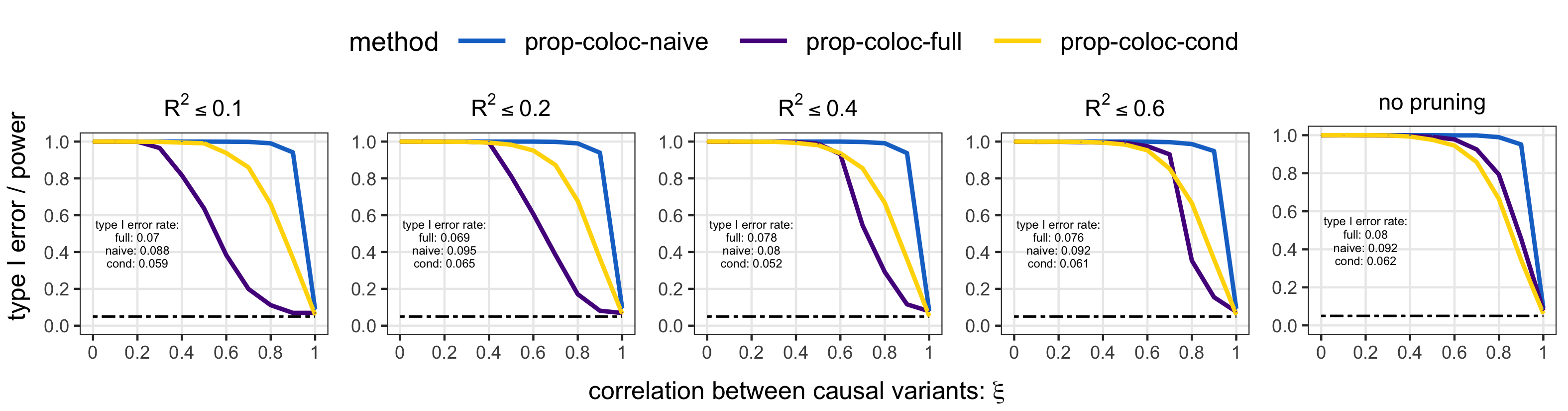}\\
\emph{\small{}Figure 5.}\emph{ }\emph{\footnotesize{}Power of proportional
colocalization tests varying with the correlation between causal variants,
under no mis-measured genetic correlations. The nominal size of the
tests was set at 0.05. }{\footnotesize\par}
\par\end{center}

Although considering only weakly correlated variants appears to control
type I error rates of the prop-coloc-full test under mis-measured
genetic correlations, this would obviously harm the power of the test.
This is verified in Figure 5, which plots the power of the prop-coloc-full
test under different pruning thresholds and under no mis-measured
genetic correlations. Stricter pruning thresholds run the risk of
omitting true causal variants from the analysis, which is the case
when the square of the correlation between causal variants $\xi^{2}$
is greater than the $R^{2}$ pruning threshold; it can be seen in
Figure 5 that the power of prop-coloc-full starts to dip at that moment.

\subsubsection{Multiple causal variants}

When there are multiple causal variants for either trait, proportional
colocalization approaches test the null hypothesis that genetic variants
proportionately colocalize, which is a specific type of colocalization
as discussed in Section 2.1.1. This section presents the power performance
of colocalization tests varying with the non-proportionality of colocalized
genetic associations. The power of the coloc method, which uses an
assumption of a single causal variant, is also shown to be affected
by the non-proportionality of colocalized signals. 

The simulation design is as described in Section 2.4.1 with the exception
of the true variant effects on traits, $\gamma_{1}$ and $\gamma_{2}$.
Two distinct causal variants $(Z_{j1},Z_{j2})$ were selected at random,
and the vector of variant effects $\gamma_{2}$ on trait $2$ was
set to be a vector of zeros apart from its $j_{1}$-th element which
was set to $0.5(1-\delta)$, and its $j_{2}$-th element which was
set to $0.5(1+\delta)$. The vector of variant effects $\gamma_{1}$
on trait $1$ was set to be a vector of zeros apart from its $j_{1}$-th
element which was set to $0.5(1+\delta)\eta_{0}$, and its $j_{2}$-th
element which was set to $0.5(1-\delta)\eta_{0}$. The non-proportionality
parameter $\delta$ was varied between $0$ and $1$; note that when
$\delta=0$, $\gamma_{1}=\gamma_{2}\eta_{0}$ so that the proportional
colocalization hypothesis holds. Then, as $\delta$ moves away from
$0$ and towards $1$, the non-proportionality of the colocalized
signals increases. 

Figure 6 presents the power of colocalization tests varying with the
non-proportionality of the causal genetic variant associations $\delta$.
The results provide further evidence that using many genetic variants
in the proportional colocalization test can inflate type I error rates
in small samples. In contrast, the conditional test was again able
to control type I error rates near the nominal $0.05$ level across
the range of sample sizes $n$ and proportionality constants $\eta_{0}$
considered. Moreover, prop-coloc-cond test was generally also more
powerful than prop-coloc-full. For example, the rejection rate of
prop-coloc-cond was around 33 percent higher than the rejection rate
of prop-coloc-full when $n=1000$, $\eta_{0}=0.5$, and $\delta=0.2$. 

Non-proportionality of colocalized genetic associations can also be
seen to affect the power of coloc, with the test favouring the hypothesis
of distinct causal variants when the non-proportionality of colocalized
signals is high. For the specific case of small samples $(n\leq1000)$
and proportionality constants of $\eta_{0}=0.1$ and $\eta_{0}=0.2$,
coloc was more powerful than coloc-SuSiE. But for all other cases,
coloc-SuSiE was generally much more powerful in detecting colocalization,
although high non-proportionality of colocalized signals also harmed
the power of the approach. 
\begin{center}
\includegraphics[width=16.5cm]{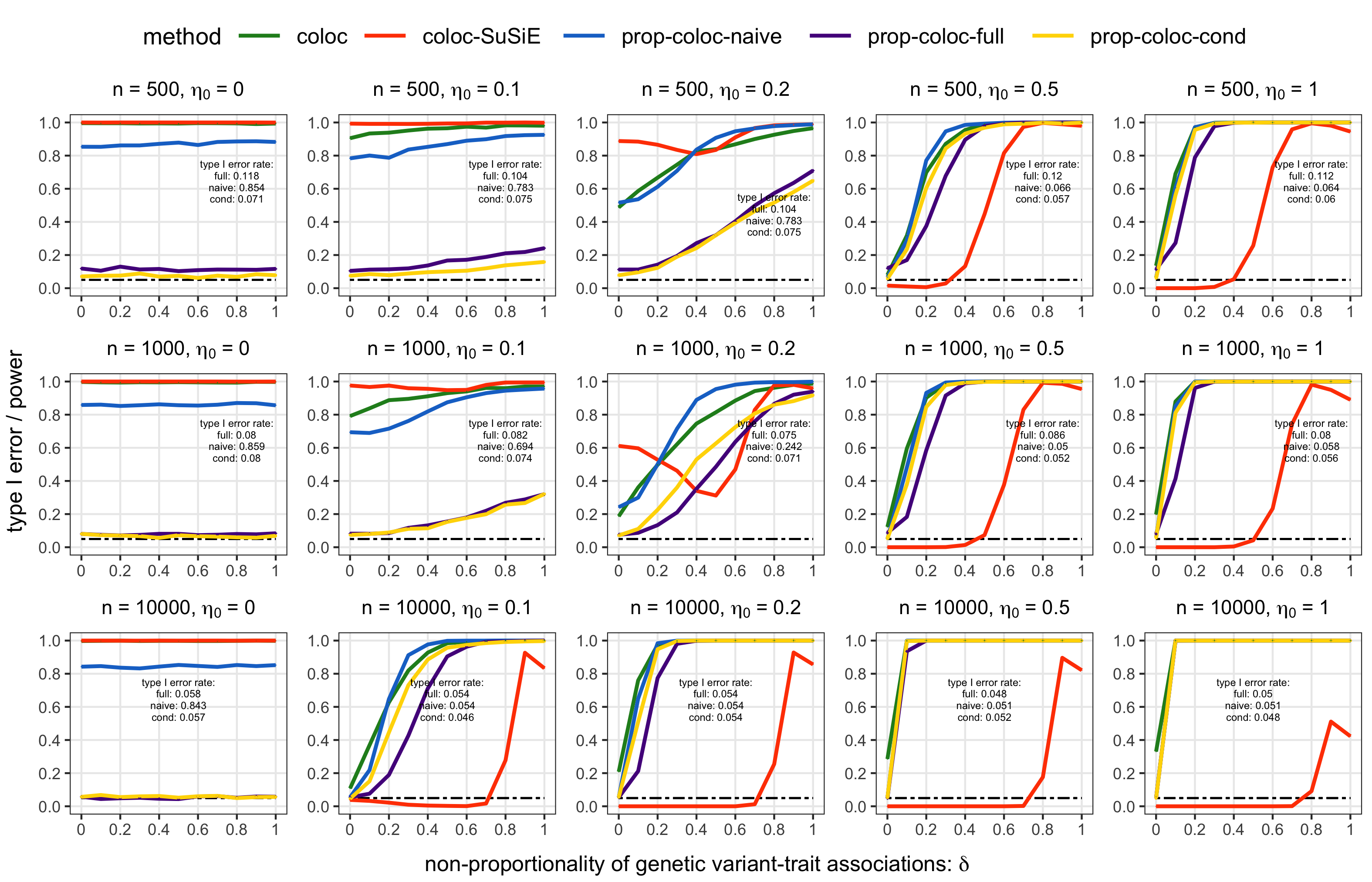}\\
\emph{\small{}Figure 6.}\emph{ }\emph{\footnotesize{}The rejection
frequencies of colocalization tests, varying with the non-proportionality
of causal genetic variant associations. The traits proportionately
colocalize only when $\delta=0$. The nominal size of proportional
colocalization tests was set at 0.05. For coloc and coloc-SuSiE the
rejection rates of the colocalization hypothesis, $P(H4<0.5)$, are
plotted.}{\footnotesize\par}
\par\end{center}

We conclude this section by summarizing our findings from the simulation
study. First, using many variants in the proportional colocalization
test can result in inflated type I error rates in small samples, but
so can naive application of the test based on the two variants most
strongly associated with traits. In contrast, the conditional proportional
colocalization test which accounts for the uncertainty in variant
selection has better finite-sample size properties, and the test can
also be more powerful when detecting non-proportionality of colocalized
signals compared to also testing the proportionality of many irrelevant
variants. 

Further, using many correlated genetic variants in proportional colocalization
tests can result in highly inflated type I error rates under mis-measured
genetic correlations. While pruning to weakly correlated variants
can alleviate type I error inflation in this regard, it also harms
the power of the test when the causal variants are more correlated.
Finally, under multiple causal variants, coloc-SuSiE is better placed
to detect colocalization under a wide range of scenarios, although
the power to detect non-proportional colocalization is not as high
in settings of small sample sizes and when the association of causal
variants with one trait are weak.

\subsection{Empirical results}

\subsubsection{Overview of results}

Figure 7 presents the results of colocalization tests. The nominal
size of the proportional colocalization tests was set at 0.05; where
the p-value of the test exceeds 0.05, we retained the null hypothesis
of proportional colocalization, and those results are marked with
a cross if in addition the p-value of the LM test is less than 0.05
(which indicates evidence for a non-zero proportionality constant).
For coloc, the posterior probability of ``H4'', the hypothesis that
the traits colocalize, is indicated. A posterior probability of H4
greater than 0.5 was taken as evidence for colocalization, and those
results are marked with a cross. The same applies for coloc-SuSiE,
but where a posterior probability of H4 greater than 0.5 at \textsl{at
least} one locus was taken as evidence for colocalization.

We first note that prop-coloc-full rejects the proportional colocalization
hypothesis in all but one pairwise trait comparison, which involves
the two heart tissues. The prop-coloc-cond test also retains the proportional
colocalization hypothesis for the heart tissues, as well as retaining
the null hypothesis in a further 25 out of the 45 pairwise trait comparisons
considered. Of these 25, only 11 pairs of traits were retained after
using the LM test to rule out the proportional colocalization hypothesis
where there is insufficient evidence for a non-zero proportionality
constant. The more conservative performance of the prop-coloc-cond
test suggests that in some pairwise comparisons, the proportionality
of variant--trait associations may hold for the two most relevant
variants, but not across other weaker associated variants.
\begin{center}
\includegraphics[width=8.5cm]{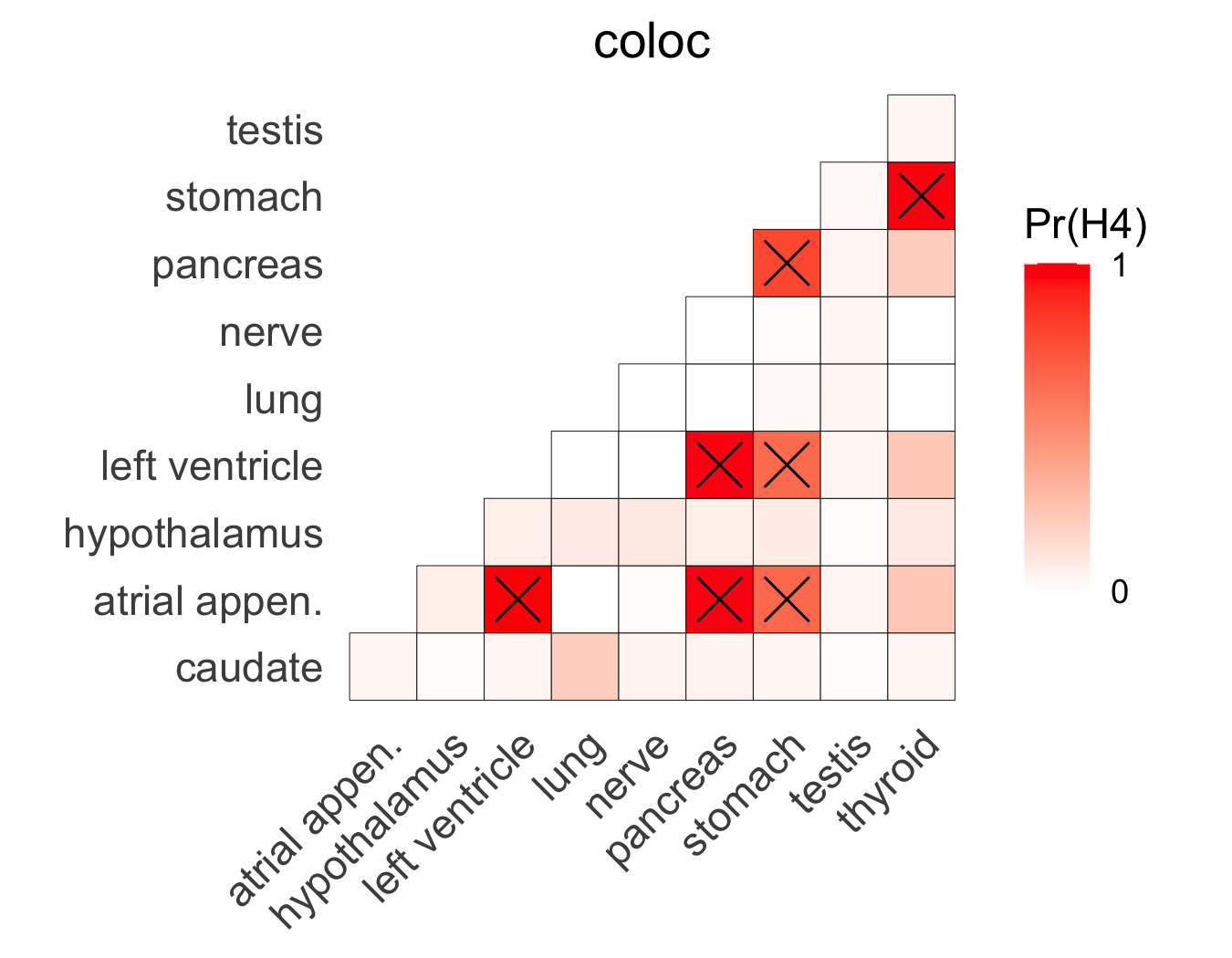}\includegraphics[width=8.5cm]{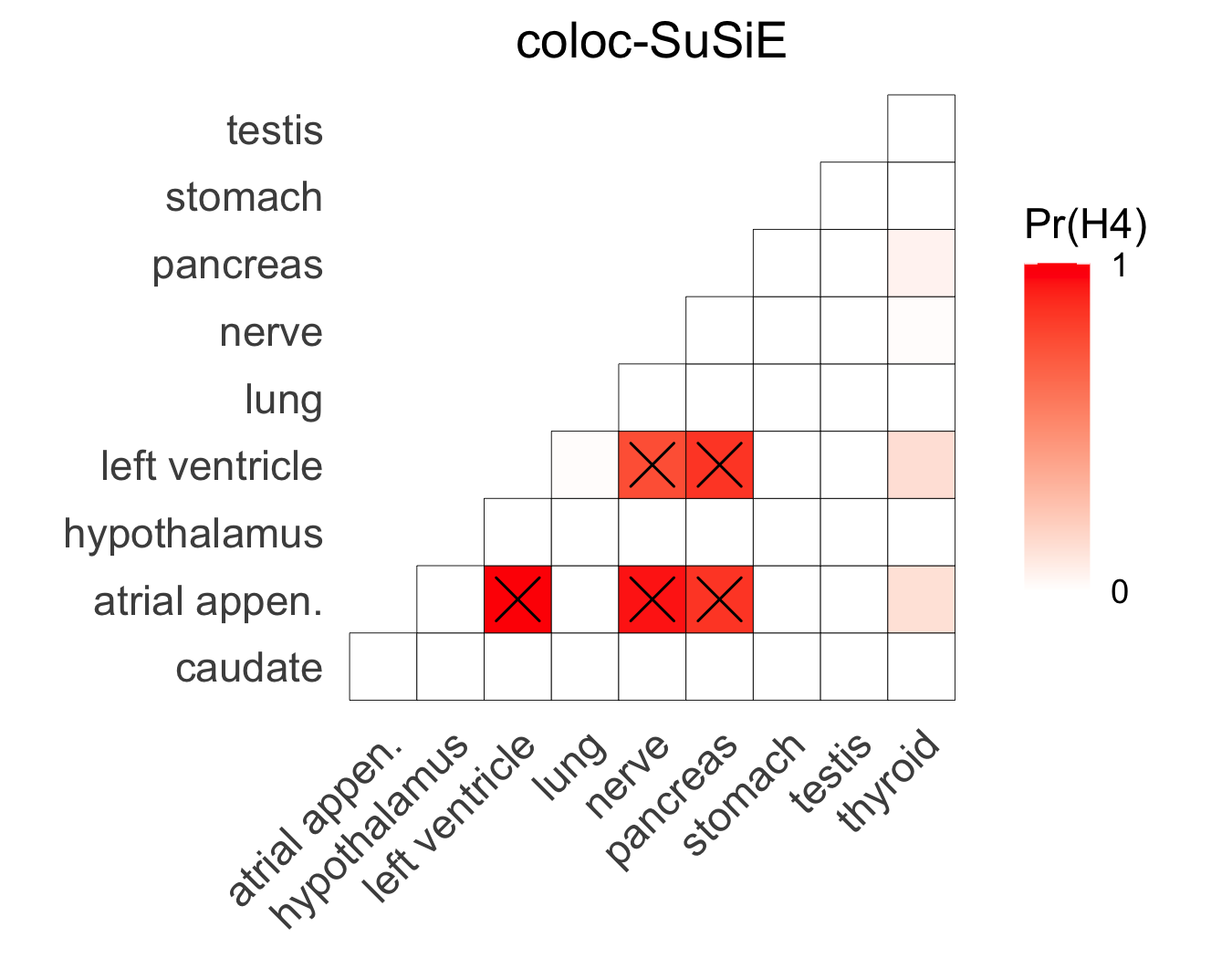}\\
\includegraphics[width=8.5cm]{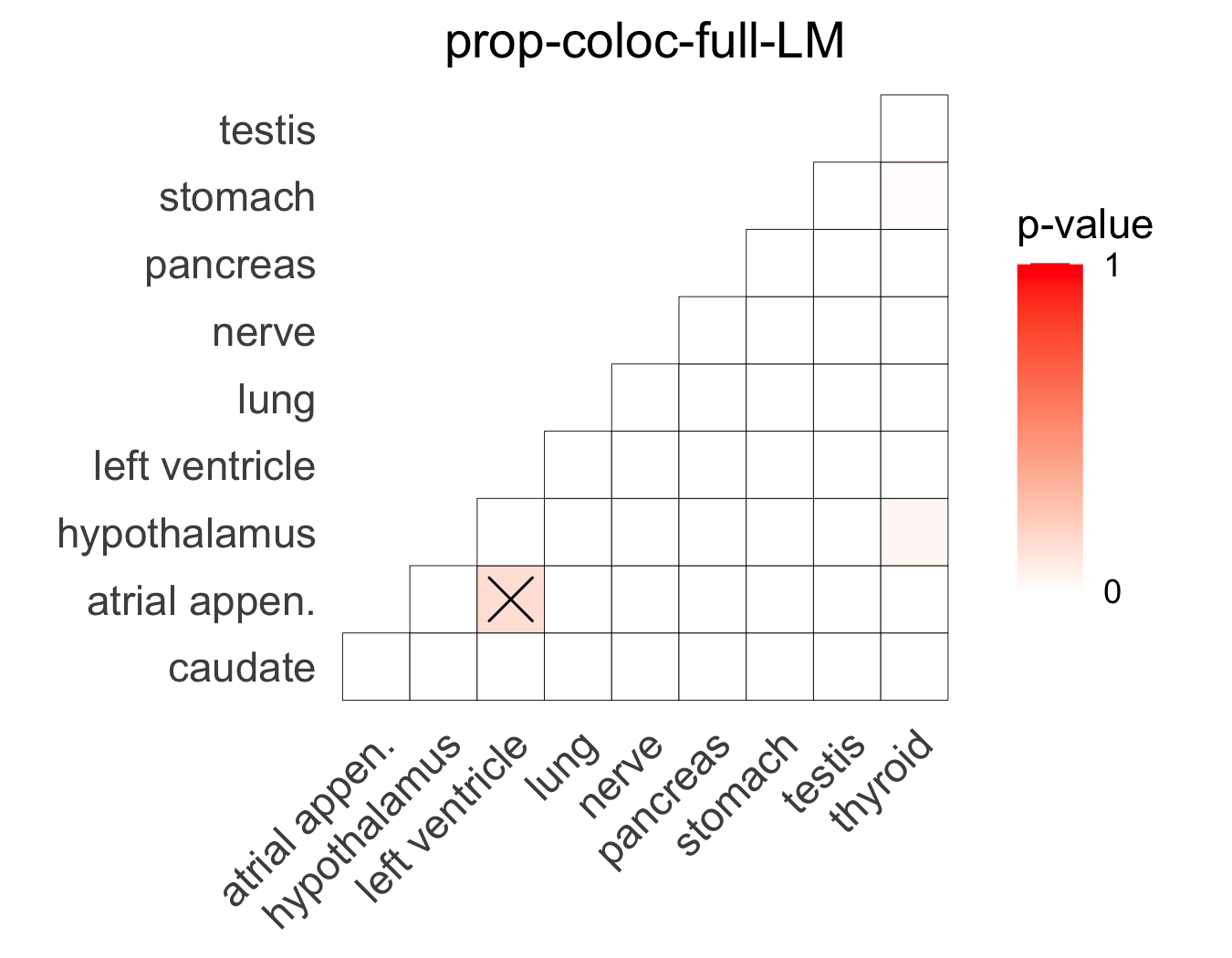}\includegraphics[width=8.5cm]{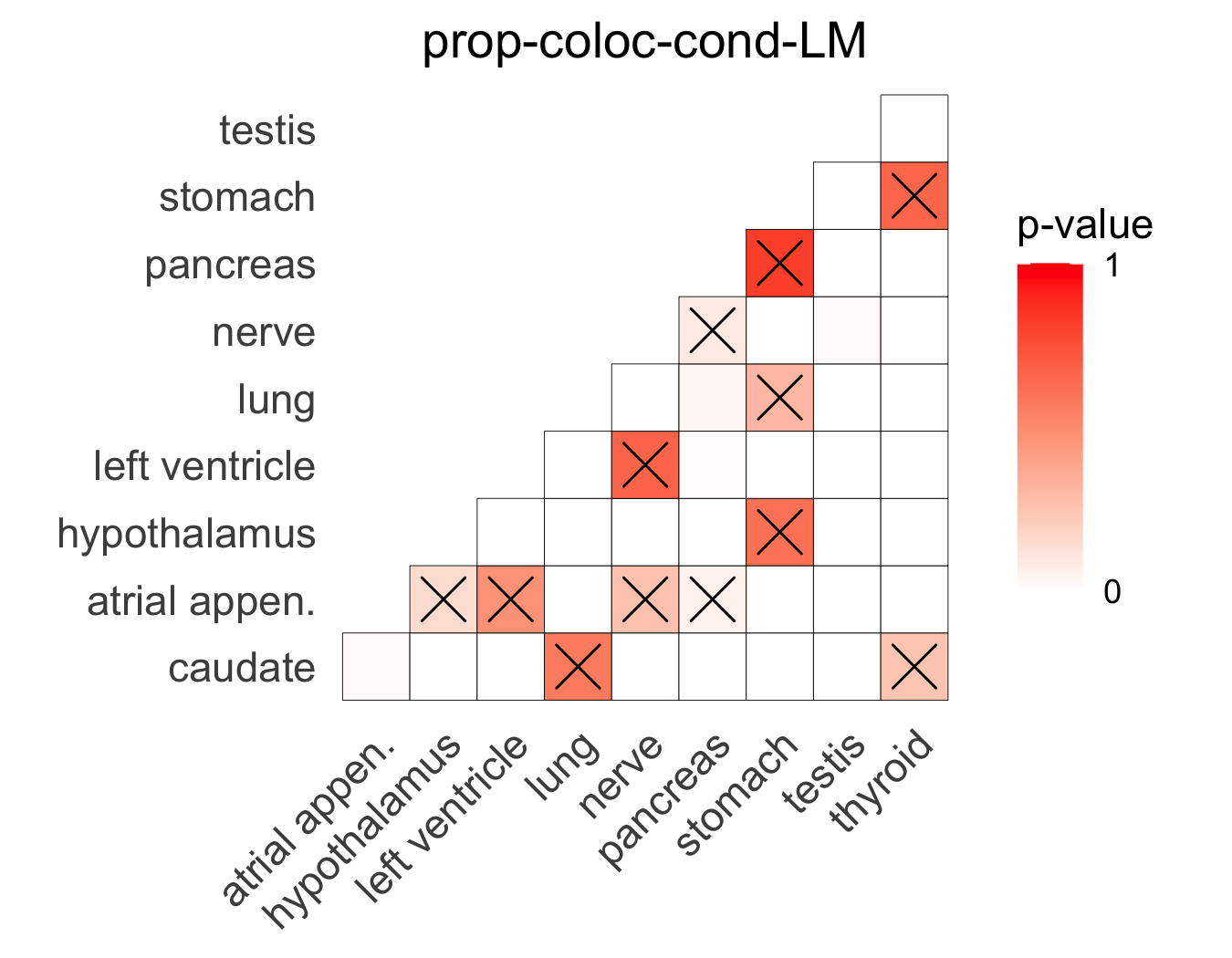}\\
\emph{\small{}Figure 7.}\emph{\footnotesize{} The p-values of proportional
colocalization tests, and posterior probabilities of colocalization
``H4'' of coloc and coloc-SuSiE tests. Where the proportional colocalization
tests do not reject the null hypothesis of proportional colocalization
(p-value > 0.05) and the LM tests do reject the null of a zero proportionality
constant (p-value < 0.05) are indicated by a cross. Similarly, where
coloc and coloc-SuSiE detect colocalization (posterior probability
of H4 > 0.5) are indicated by a cross.}{\footnotesize\par}
\par\end{center}

Of the 5 pairwise trait comparisons judged to colocalize by the coloc-SuSiE
method, 4 of the 5 are supported with the results of prop-coloc-cond,
3 of the 5 are consistent with the results of coloc, and only the
colocalization finding for the two heart tissues is supported by the
results of all other methods. Of the 7 colocalization findings by
the coloc method, 3 are supported by the results of coloc-SuSiE, and
4 are supported by the results of prop-coloc-cond.

\subsubsection{Differing results across colocalization methods}

The results in Figure 7 show that in 30 out of 45 pairwise trait comparisons,
all four methods that we considered (prop-coloc-cond, prop-coloc-full,
coloc, and coloc-SuSiE) reject a colocalization hypothesis. For the
remaining pairwise trait comparisons, there is some disagreement in
the colocalization evidence provided. As discussed in Section 2.3,
this is primarily because each of the four methods are testing different
hypotheses. We now discuss specific cases where the methods suggest
similar and contrasting evidence. 

For each pairwise trait comparison in Figures 8--14, we plot genetic
associations with the two traits (gene expression in two tissues)
based on univariable ($X_{k}$ on $Z_{j}$) and multivariable ($X_{k}$
on $Z=(Z_{1},\ldots,Z_{J})^{\prime}$) linear regressions. The two
lead variants selected by the prop-coloc-cond method are circled in
green. The slopes in the rightmost plots indicate the estimated proportionality
constant from the prop-coloc-full (red) and prop-coloc-cond (green)
methods. In the two leftmost plots, causal variants detected by coloc-SuSiE
are highlighted. The plots for the remaining pairwise trait comparisons
not discussed here are provided in Supplementary Information.

\subsubsection{All methods conclude evidence for colocalization}

One unanimous finding by all proportional and enumeration colocalization
methods was the colocalization of gene expression in the two heart
tissues (atrial appendage and left ventricle).
\begin{center}
\includegraphics[width=16.5cm]{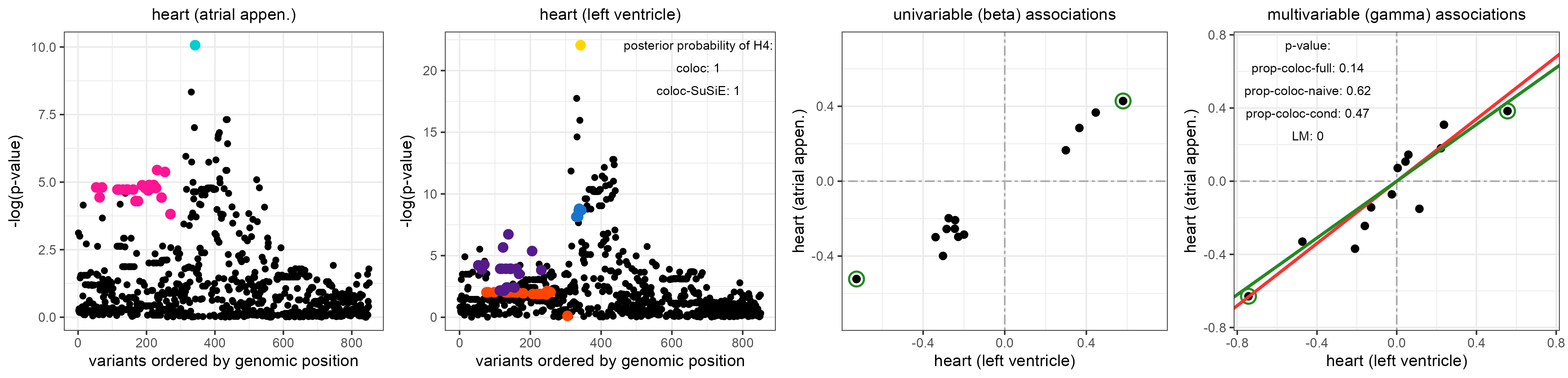}\\
\emph{\small{}Figure 8.}\emph{\footnotesize{} Colocalization results
for gene expression in two heart tissues. Further details for the
plots are discussed at the start of Section 3.2.2. }{\footnotesize\par}
\par\end{center}

Figure 8 suggests evidence of at least two causal variants that are
shared by both traits. Moreover, the genetic associations with each
trait appear to be proportional; the proportionality constant was
estimated to be 0.850 (LM p-value 0.000) in the prop-coloc-full test,
and 0.775 (LM p-value 0.000) in the prop-coloc-cond test. The posterior
probability of H4 (colocalization) was greater than 0.999 using the
coloc method. The coloc-SuSiE method suggested colocalization with
posterior probability greater than 0.999 at one variant, and greater
than 0.804 at another variant. In this case, we would conclude that
\textsl{GLP1R} gene expression in the atrial appendage and left ventricle
tissues colocalize. 

\subsubsection{Only coloc and coloc-SuSiE conclude evidence for colocalization}
\begin{center}
\includegraphics[width=16.5cm]{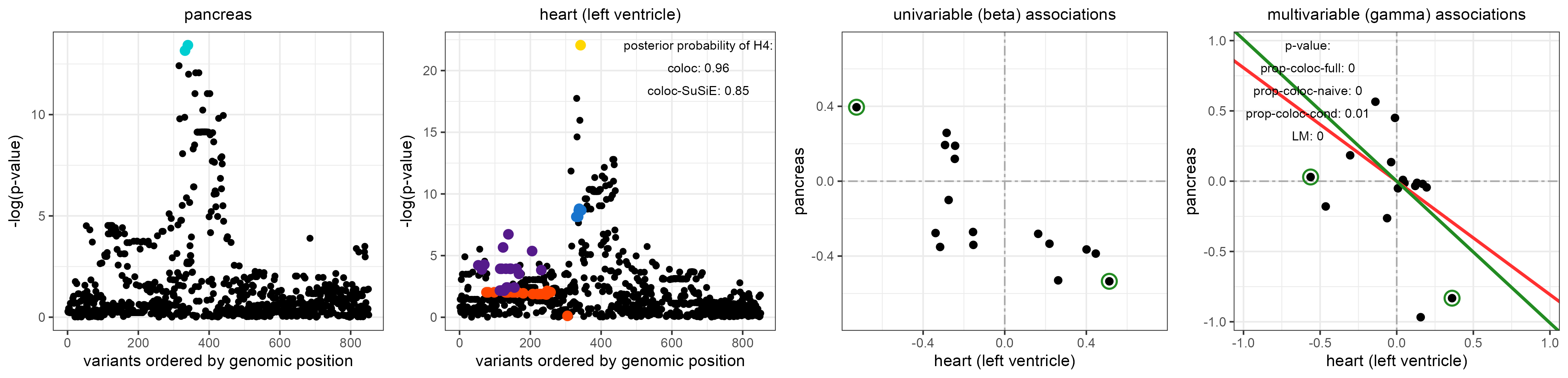}\\
\emph{\small{}Figure 9.}\emph{\footnotesize{} Colocalization results
for gene expression in pancreas and heart (left ventricle) tissues.
Further details for the plots are discussed at the start of Section
3.2.2. }{\footnotesize\par}
\par\end{center}

Only the coloc and coloc-SuSiE methods suggested colocalization evidence
for gene expression in the pancreas and heart (left ventricle) tissues.
In Figure 9, there appears to be one causal variant shared by gene
expression in the pancreas and heart (left ventricle) tissues. 

The coloc-SuSiE method indicates there may be only one causal variant
for pancreas, and multiple causal variants for left ventricle. In
this case, we would not expect genetic associations with each trait
to proportional. Indeed, there appears to be considerable heterogeneity
in genetic associations, leading to a rejection of the proportional
colocalization hypothesis using all variants. Further, the two circled
lead variants suggest very different estimates of the proportionality
constant, which also leads to a rejection of the proportional colocalization
hypothesis under the prop-coloc-cond method. Overall, we conclude
that there is a causal variant shared by \textsl{GLP1R} gene expression
in the pancreas and left ventricle tissues, but the proportional colocalization
hypothesis does not hold. 

\subsubsection{Only coloc and prop-coloc-cond conclude evidence for colocalization}
\begin{center}
\includegraphics[width=16.5cm]{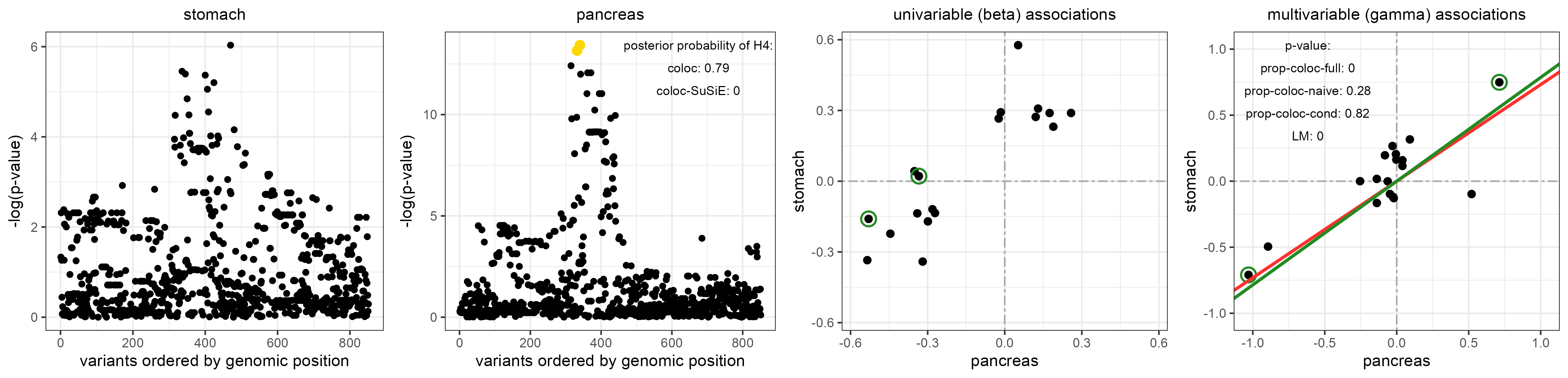}\\
\emph{\small{}Figure 10.}\emph{\footnotesize{} Colocalization results
for gene expression in stomach and pancreas tissues. Further details
for the plots are discussed at the start of Section 3.2.2. }{\footnotesize\par}
\par\end{center}

Only the coloc and prop-coloc-cond methods suggested colocalization
evidence for gene expression in the stomach and pancreas tissues.
Here, there appears to be a strong causal variant for pancreas that
is less strongly associated with stomach. Given this, coloc concludes
colocalization evidence, whereas coloc-SuSiE suggests no causal variants
for stomach. 

In the two rightmost plots of Figure 10, there is significant heterogeneity
in variant--trait associations when considering all variants, so
that the prop-coloc-full rejects the null hypothesis of proportional
colocalization. Although the two lead variants are in different quadrants,
they suggest similar estimates of the proportionality constant, and
so the prop-coloc-cond method does not reject the proportional colocalization
hypothesis (proportionality constant estimate: 0.774; LM p-value 0.002).
The result of prop-coloc-cond therefore tallies with coloc in concluding
evidence for a shared causal variant for \textsl{GLP1R} gene expression
in the stomach and pancreas tissues. 

\subsubsection{Only coloc-SuSiE and prop-coloc-cond conclude evidence for colocalization}

Only the coloc-SuSiE and prop-coloc-cond methods suggested colocalization
evidence for gene expression in the nerve and heart tissues. From
Figure 11, coloc-SuSiE detects multiple causal variants for left ventricle,
but the top signals for left ventricle and nerve appear to be different
variants. Therefore coloc does not conclude evidence for colocalization,
whereas coloc-SuSiE does because the causal variant for nerve is judged
to also be causal for left ventricle even though it is not the strongest
signal for left ventricle. 
\begin{center}
\includegraphics[width=16.5cm]{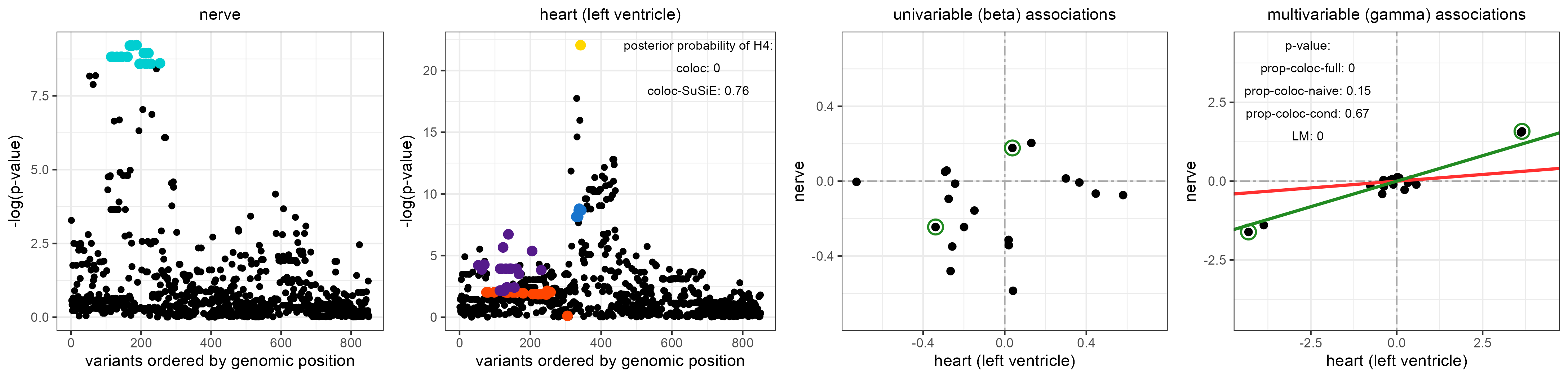}\\
\emph{\small{}Figure 11.}\emph{\footnotesize{} Colocalization results
for gene expression in nerve and heart (left ventricle) tissues. Further
details for the plots are discussed at the start of Section 3.2.2. }{\footnotesize\par}
\par\end{center}

The variant--trait associations in the two rightmost plots in Figure
11 suggest significant heterogeneity when considering all variants,
and hence prop-coloc-full rejects the proportional colocalization
hypothesis. However, the two lead variants suggest quite similar estimates
of a positive proportionality constant (estimate: 0.324; LM p-value:
0.003), and thus prop-coloc-cond retains the null of proportional
colocalization. A clear conclusion is difficult; the top signals for
the two traits appear to be distinct, but the lead variant for nerve
may be causal for both traits. 

\subsubsection{Only coloc concludes evidence for colocalization}
\begin{center}
\includegraphics[width=16.5cm]{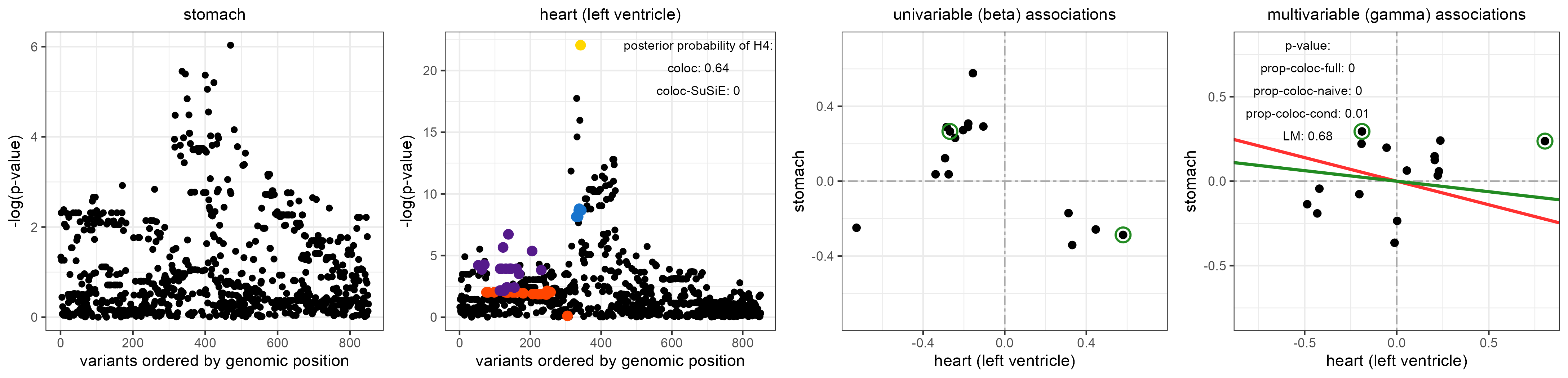}\\
\emph{\small{}Figure 12.}\emph{\footnotesize{} Colocalization results
for gene expression in stomach and heart (left ventricle) tissues.
Further details for the plots are discussed at the start of Section
3.2.2. }{\footnotesize\par}
\par\end{center}

Only the coloc method suggested colocalization evidence for gene expression
in the stomach and heart tissues. Similar to Figure 10, a top signal
in Figure 12 for one of the traits (in this case, left ventricle)
appears to be located near to the top, but weaker, signal for the
other trait (stomach) leading coloc to favour a colocalization hypothesis,
whereas coloc-SuSiE concludes there is no causal variant for stomach. 

There appears to be significant heterogeneity in variant--trait associations,
leading to a rejection of the proportional colocalization hypothesis
from the prop-coloc-full method. Moreoever, in the rightmost plot
in Figure 12, the two lead variants are in different quadrants and
suggest no coherent estimate of the proportionality constant, and
the LM test cannot reject the null hypothesis of a zero proportionality
constant. This could represent a setting similar to model 1 in Section
2.3 where only coloc tends to detect colocalization, and where the
linear model of proportional colocalization does not hold between
the two lead variants. 

\subsubsection{Only prop-coloc-cond concludes evidence for colocalization}
\begin{center}
\includegraphics[width=16.5cm]{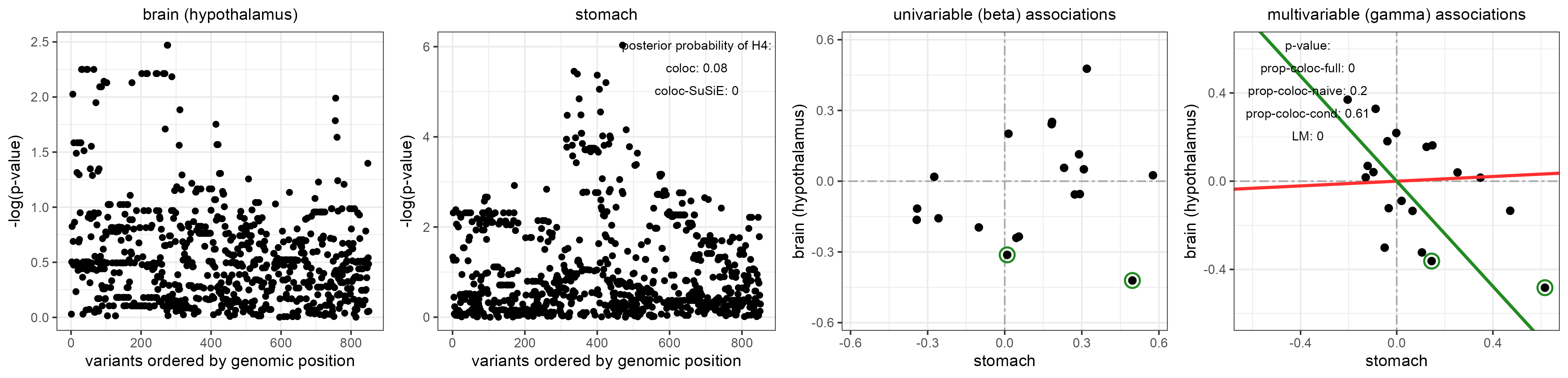}\\
\emph{\small{}Figure 13.}\emph{\footnotesize{} Colocalization results
for gene expression in brain (hypothalamus) and stomach tissues. Further
details for the plots are discussed at the start of Section 3.2.2. }{\footnotesize\par}
\par\end{center}

Only the prop-coloc-cond method suggested colocalization evidence
for \textsl{GLP1R} gene expression in the brain (hypothalamus) and
stomach tissues. In Figure 13, the top signal for brain (hypothalamus)
is quite weak, with coloc favouring a ``H2'' hypothesis of a causal
variant only for stomach (with posterior probability 0.680). The coloc-SuSiE
method does not detect a causal variant for either trait. 

Again, the heterogeneity across all variant--trait associations leads
the prop-coloc-full test to reject the null hypothesis of proportional
colocalization, but the two lead variants suggest a similar value
of the proportionality constant (estimate: -1.191; LM p-value: 0.003),
so that the prop-coloc-cond test retains the null hypothesis. 

\subsubsection{None of the methods conclude evidence for colocalization}

Finally, we note a case where a colocalization hypothesis was not
supported by any of the methods considered. From the leftmost manhattan
plots in Figure 14, there is no clear indication of a shared causal
variant between testis and stomach. In this case, coloc favours a
``H2'' hypothesis of a causal variant only for stomach, whereas
coloc-SuSiE does not detect a causal variant for either trait. 
\begin{center}
\includegraphics[width=16.5cm]{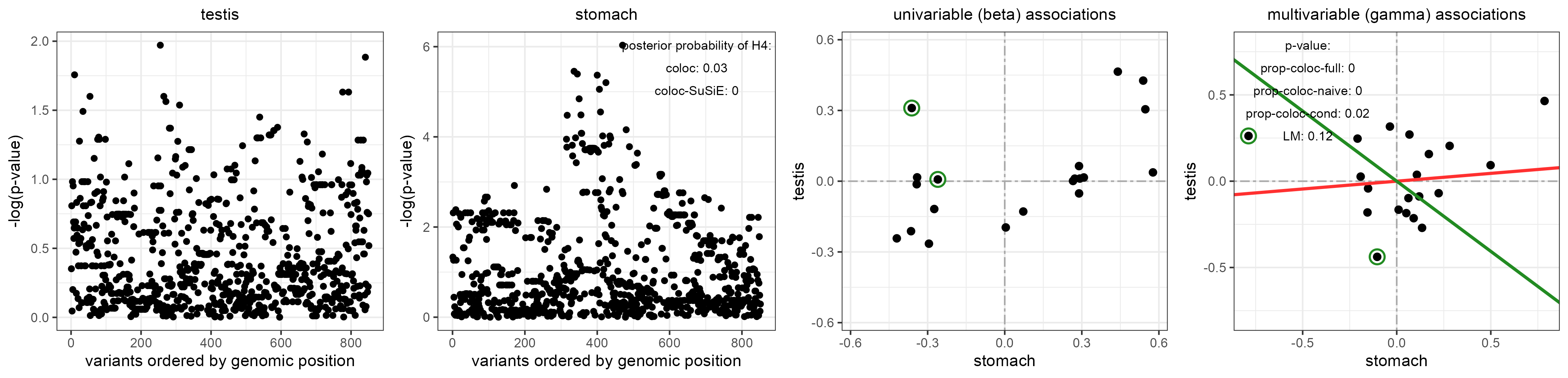}\\
\emph{\small{}Figure 14.}\emph{\footnotesize{} Colocalization results
for gene expression in testis and stomach tissues. Further details
for the plots are discussed at the start of Section 3.2.2. }{\footnotesize\par}
\par\end{center}

The two rightmost plots of variant-trait associations show no clear
linear trend; there is excessive heterogeneity which leads to a rejection
of the proportional colocalization hypothesis under both the prop-coloc-full
and prop-coloc-cond methods. Further, the LM test is unable to suggest
evidence for a non-zero proportionality constant, indicating there
is no causal variant for testis. Overall, there is unlikely to be
a causal variant for testis, but there may be a causal variant for
stomach. 

\section{Discussion}

Enumeration and proportional colocalization approaches differ in both
the assumptions they use, and the null hypotheses or priors that are
maintained in the absence of evidence to the contrary. Therefore we
may be more confident in colocalization evidence that is supported
by different methods in the spirit of triangulation \citep{Munafo2018}.
Where the methods give different answers, careful interrogation can
give insight into the most likely causal model, as in the examples
discussed in Section 3.2.4--3.2.8. 

In this work, we primarily focused on the proportional colocalization
approach, and derived a conditional test that aims to account for
uncertainty in variant selection, in order to overcome the issue of
inflated type I error rates highlighted in \citet{Wallace2013}. Our
simulation evidence suggests that the conditional test has competitive
finite-sample size properties compared with a proportional colocalization
test that compares proportionality of variant--trait associations
for a large number of variants. 

Specific cases where we may expect contrasting evidence include: (i)
when there is a single causal variant for one trait that is weakly
associated with another trait, and other variants that more weakly
associated with only one trait -- here, only coloc tends to provide
evidence for colocalization (Model 1 of Figure 2 in Section 2.3, and
Figure 12 in Section 3.2.7); (ii) when there is non-proportional colocalization
at multiple causal variants -- here, only coloc-SuSiE tends to provide
evidence for colocalization (Model 2 of Figure 2 in Section 2.3);
(iii) under small sample sizes, and relatedly, when proportionality
of variant--trait associations holds for weak causal variants, the
conditional proportional colocalization test tends not to reject the
proportional colocalization hypothesis. 

Reasons for the popularity of the coloc method are clear: it can be
implemented straightforwardly with minimal data requirements, needing
only summarized data on the traits of interest, and it gives clear
output that provides direct evidence on the probability of colocalization.
However, the method also has weaknesses: it is often sensitive to
specification of the priors (particularly for $p_{12}$, the prior
probability of a variant being causal for both traits), it assumes
a single causal variant for both traits, it focuses on the lead signals
for both traits, and it can provide ambiguous conclusions with no
strong evidence either for or against colocalization \citep{Burgess2023}.
The coloc-SuSiE method addresses some of these issues, allowing multiple
causal variants and hence reducing the focus on the lead signals. 

The proportional colocalization method operates in a different paradigm,
complicating direct comparison of the methods. Both coloc and coloc-SuSiE
operate in a Bayesian paradigm, whereas proportional colocalization
is implemented in a frequentist paradigm. This means that proportional
colocalization can either reject the null hypothesis of colocalization,
or not reject this hypothesis: the latter could represent lack of
strong evidence against colocalization rather than strong evidence
to support colocalization. In cases where genetic associations with
one trait are not strong, proportional colocalization could therefore
act as a negative filter, triaging out situations where colocalization
is not supported by the data. If coloc/coloc-SuSiE and proportional
colocalization give concordant results, this can be interpreted as
stronger evidence than provided by either method individually. If
they give discordant results, a variety of explanations are possible:
this could reflect weak evidence both for and against colocalization,
different definitions of colocalization tested by the approaches,
or a focus on lead variants (in coloc) versus consideration of several
variants (particularly in prop-coloc-full). 

Biology is complex and messy. In proposing this test for proportional
colocalization, we provide another method that tests for colocalization
that may agree or disagree with the commonly-used coloc method. Where
it agrees, this provides the analyst with stronger evidence supporting
or refuting colocalization; where it disagrees, this provides a caution
that the results we see are not black-and-white, but reflect the complexity
in the biological mechanisms that underlie the association estimates
in our statistical models.

\subsection*{Acknowledgements}

We thank Chris Wallace and Amy Mason for helpful discussions. This
research was funded by the United Kingdom Research and Innovation
Medical Research Council (MC-UU-00002/7 and MC-UU-00002/18), and supported
by the National Institute for Health Research Cambridge Biomedical
Research Centre (BRC-1215-20014). S.B. is supported by the Wellcome
Trust (225790/Z/22/Z). 

\subsection*{Data availability}

The summary genetic association data used for the analyses described
in this manuscript were obtained from the Genotype-Tissue Expression
(GTEx) Portal (project version 8) at https://gtexportal.org/home GTEx,
and UK Biobank Linkage Disequilibrium Matrices were obtained from
the AWS Open Data Sponsorship Program at\\
https://registry.opendata.aws/ukbb-ld.

\subsection*{Code availability}

An R package to apply the methods proposed in this manuscript is available
at\\
https://github.com/ash-res/prop-coloc, along with R code that reproduces
the empirical results in this manuscript.

\subsection*{Declaration of interests}

J.C.W. is a member of scientific advisory boards/consultancy for Relation
Therapeutics and Silence Therapeutics, and acknowledges ownership
of GlaxoSmithKline shares. 

\bibliographystyle{chicago}
\bibliography{propcoloc}

\end{document}